\documentclass[prd,aps,amssymb,amsmath,tightenlines,nofootinbib]{revtex4}
\usepackage{graphicx}
\usepackage{slashed}
\usepackage[colorinlistoftodos]{todonotes}
\usepackage{appendix}
\newcommand{\tmop}[1]{\ensuremath{\operatorname{#1}}}
\newcommand{\be}{\begin{equation}}
\newcommand{\ee}{\end{equation}}

\def\cPT{$\mathcal P \mathcal T~$}
\def\eq{\eqref}

\begin{document}

\preprint[{\leftline{KCL-PH-TH/2022-{\bf 05}}

\title{Schwinger-Dyson equations and mass generation for an axion theory with a \cPT symmetric   Yukawa fermion interaction}

\author{N. E. Mavromatos$^{a, b}$}
\author{Sarben Sarkar$^b$}
\author{A Soto$^{c
}$}

\affiliation{$^a$Physics Department, School of Applied Mathematical and Physical Sciences, National Technical University of Athens , Athens 157 80, Greece\\
$^b$Theoretical Particle Physics and Cosmology Group, Department of Physics, King's~College~London, Strand, London WC2R 2LS, UK\\
$^c$School of Mathematics, Statistics and Physics, Newcastle University, Newcastle upon Tyne, NE1 7RU, UK\\
}

\begin{abstract} 
A nonperturbative Schwinger-Dyson analysis of mass generation is presented for a non-Hermitian \cPT-symmetric field theory in four dimensions of an axion coupled to a Dirac fermion.The model is motivated by phenomenological considerations.The axion has a quartic self-coupling $\lambda$  and a Yukawa coupling $g$ to the fermion.  The  Schwinger-Dyson equations are derived for the model with generic couplings. In the non-Hermitian case there is an additional nonperturbative contribution to the scalar mass. In a simplified rainbow analysis the solutions for the SD equations, are given for different regimes of the couplings $g$ and $\lambda$.

\end{abstract}


\maketitle
\section{Introduction}
\label{sec1}
 The emergence a new class of non-Hermitian quantum-mechanical Hamiltonians with a discrete \cPT symmetry~\cite{R2}, where $\mathcal{P}$ is a linear operator and $\mathcal{T}$ is an antilinear operator, is inspiring the development of several non-Hermitian field theories~\cite{R1,R10,R6a,R5a} in the search for descriptions of physics beyond the Standard Model (SM) of particle physics.  
 \cPT-symmetric quantum mechanical theories possess real energy eigenvalues. For a class of   \cPT-symmetric Hamiltonians, \emph{all}  the energies is rigorously shown to be  real~\cite{R2a,R2b}.~\cPT-symmetric theories belong to the general class of unitary  pseudo-Hermitian~\cite{R3} quantum theories where the inner product on the Hilbert space is different from the conventional Dirac inner product~\cite{R3a}. 

 One way of formulating quantum mechanics is through path integrals.~Going from \cPT-symmetric quantum mechanics to quantum field theory raises additional complications in the description of path integrals such as renormalisation. However, in order to construct viable fundamental theories based on \cPT symmetry, it is necessary to construct path integrals in complex field space which are formally convergent.
  In a recent paper~\cite{R4}, we formulate path integrals for such (non-gauge) field theories. Path integral quantisation has the advantage that Green's functions can be calculated without an explicit construction of the inner product on the Hilbert space~\cite{R4a}. The understanding of \cPT~symmetric field theory cannot rely on a purely perturbative treatment.
  This is not hard to appreciate: an upside-down quartic scalar potential, which is \cPT-symmetric, is conventionally unstable. This instability is due to tunnelling which is a nonperturbative phenomenon.~\cPT symmetry introduces a nonperturbative effect which tames this instability and leads to a stable vacuum~\cite{R2}. 
  
  We study in this paper the \emph{nonperturbative} phenomenon of dynamical mass generation using Schwinger-Dyson (SD) equations for a \cPT-symmetric quantum field theory motivated by gravitational axion physics. In low spacetime dimensions~\cPT symmetry allows an alternative way to generate nonperturbatively a scalar mass when the bare Lagrangian has no mass.
The model that we study is described by the Lagrangian (in $D=3+1(=4)$ Minkowski spacetime dimensions,  with metric signature $(+, -, -, -)$ 
):\footnote{In earlier papers~\cite{R10,R6a,R5a} a related  model, with both Hermitian and non-Hermitian Yukawa interactions, is studied using SD equations for dynamical mass generation  but without the quartic scalar interaction which is important in a renormalisable \cPT-symmetric field theory~\cite{R4}.}
 \begin{equation}
\label{E1}
\mathcal{L}= \frac{1}{2} \partial_{\mu} \phi \partial^{\mu} \phi -
\frac{M^2}{2} \phi^2 + \bar{\psi} \left( i \slashed{\partial} - m\right)
\psi - i g \bar{\psi} \gamma^5 \psi \phi - \frac{\lambda}{4!} \phi^{2} \left( i\phi \right)^{\delta } 
=L_{B} +L_{F},
\end{equation}
where 
\be\label{E1a}
L_{B}=\frac{1}{2} \partial_{\mu} \phi \partial^{\mu} \phi -
\frac{M^2}{2} \phi^2 - \frac{\lambda}{4!} \phi^{2} \left( i\phi \right)^{\delta } ,
\ee

\be\label{E1b}
L_{F}= \bar{\psi} \left( i \slashed{\partial} - m\right)\psi - i g \bar{\psi} \gamma^5 \psi \phi,
\ee
$\phi$ denotes a pseudoscalar (axion) field, $\psi$ denotes a generic Dirac field and $\delta$ is a real parameter. 
In the Dirac representation of gamma matrices, the conventional discrete  transformations on $\psi$~\cite{RR5} are:
\begin{align}\label{E1c}
\mathcal{P}\psi (t,\vec{x})\mathcal{P^{\it{-1}}}=\gamma^{0}\psi (t,-\vec{x}), \quad \mathcal{T}\psi (t,\vec{x})\mathcal{T^{\it{-1}}}=i\gamma^{1}\gamma^{3}\psi(-t,\vec{x}), \quad {\mathbb C} \psi \left(t,\overrightarrow{x}\right) {\mathbb C}^{-1}=i\gamma^{2}\psi^{\dagger}\left(t,\overrightarrow{x}\right).
\end{align}
where $\mathbb C$ denotes the  charge conjugation operator~\cite{RR5}  and $\mathcal{T}$ is the \emph{antilinear} time-reversal operator. Also,  in the case of \cPT-symmetric non-Hermitian theory, under the action of $\mathcal P$ and $\mathcal T$,  the charge-conjugation even pseudoscalar field $\phi\left(t,\overrightarrow{x}\right)$ transforms as~\cite{R4}~\footnote{ Note that in the Hermitian case the pseudoscalar changes sign under $\mathcal{T}$~\cite{RR5}, in contrast to the postulated~\cite{R4} transformation \eqref{E1d} in the \cPT non-Hermitian case.}
\begin{align}\label{E1d}
\mathcal{P}\phi (t,\vec{x})\mathcal{P^{\it{-1}}}= -\phi (t,-\vec{x}), \qquad   \mathcal{T}\phi (t,\vec{x})\mathcal{T^{\it{-1}}}= \phi (-t,\vec{x}).
\end{align}}

The self-interaction is parametrised in a way which \emph{emphasises} \cPT symmetry. When \underline{$\delta=2$ and $\lambda>0$} we have a quartic self-interaction which represents a potential $V(\phi)$ which is \underline{unbounded below}, i.e. an upside-down potential. In the Appendix we discuss the path integral quantisation of \cPT-symmetric field theories~\cite{R4}.

 The outline of our paper is as follows:
in section \ref{sec:nhystring} a physical (microscopic) motivation~\cite{Mav2020} for the above Lagrangian $\mathcal{L}$ is presented. 
In section \ref{e1b} we discuss the interplay of Hermiticity and non-Hermiticity implied by perturbative renormalisation group analysis.
 In section \ref{sec:SD} we present a non-perturbative  SD analysis which takes into account the special features of non-Hermiticity in the scalar sector. Although in low dimensions, such as $D=1$, the special features are important, for $D=4$, we will argue that a conventional approach suffices.
Solutions are derived for the SD equations when the couplings can be both Hermitian and non-Hermitian. The solution for mass generation is first derived in the rainbow approximation, in the neighbourhood of the trivial fixed point for the Yukawa and self-interaction couplings in section \ref{sec:rainbow}; an extension, beyond the rainbow approximation, where the effects of wave-function renormalization are taken into account, is performed in section \ref{sec:beyondrainbow}. This analysis yields results consistent with the solutions using the rainbow approximation but also points out the possibility of the existence of a critical Yukawa coupling, above which new types of mass generation in the non-perturbative regime of the theory might arise.
Finally,  the concluding section \ref{sec:concl} will contain a summary of the results and new directions for further investigation.
Some technical aspects of our approach are presented in two Appendices. Appendix \ref{sec:appA} contains a brief review of the most important points concerning  \cPT symmetry in bosonic path integrals, and also fermionic path integrals in subsection \ref{sec:fermionic}. 
Appendix \ref{sec:appC} deals with a generic derivation of Schwinger-Dyson equations, in $D=4$, for our Yukawa theory with (pseudoscalar) self-interactions, in both the Hermitian and non-Hermitian cases.

\section{A Microscopic model for non-hermitian Yukawa interactions \label{sec:nhystring}}

 In superstring theory~\cite{R6}, after compactification to four spacetime dimensions, the bosonic ground state of the closed string sector consists of massless fields of the gravitational multiplet. These massless fields are a scalar spin $0$ dilaton $\Phi$, the graviton $g_{\mu \nu }$ and a spin 1 antisymmetric tensor gauge field $B_{\mu \nu }\left( x\right)$  known as the Kalb-Ramond (KR) field. We will consider solutions with $\Phi=\Phi_{0}$ a constant.\footnote{The string coupling $g_{s}=\exp \left( \Phi_{0} \right)$.} The KR-field strength of the $B_{\mu \nu }\left( x\right)$ field is 
\begin{equation}
\label{E2}
\mathcal{H}_{\mu \nu \rho }\left( x\right)  =\partial_{[\mu } B_{\nu \rho ]},
\end{equation}
and to, lowest order in the string Regge slope $\alpha'$, the Euclidean effective action of the closed string bosonic sector is

\begin{equation}
\label{E3}
S_{B}=-\int d^{4}x\  \sqrt{-g} \left( \frac{1}{2\kappa^{2} } R+\frac{1}{6} \mathcal{H}_{\lambda \mu \nu }\mathcal{H}^{\lambda \mu \nu }+\ldots \right)  
\end{equation}
where $\kappa =\frac{\sqrt{8\pi } }{M_{P}}$, $M_{P}$ is the Planck mass, $g$ is the determinant of $g_{\mu\nu}$ and $R$ is the Einstein Ricci scalar. $S_{B}$ can be interpreted geometrically on noting that the KR field strength term $\mathcal{H}^2$ can be absorbed into a modified Christoffel symbol with $\mathcal H$ torsion~\cite{R5}
\begin{equation}
\label{E4}
\bar{\Gamma }^{\rho }_{\  \mu \nu } =\Gamma^{\rho }_{\  \mu \nu } +\frac{\kappa }{\sqrt{3} } \mathcal{H}^{\rho }_{\  \mu \nu }\neq \bar{\Gamma }^{\rho }_{\  \nu \mu } . 
\end{equation}
The lack of symmetry of $\bar{\Gamma }^{\rho }_{\  \mu \nu }$ is due to the antisymmetry of $\mathcal{H}^{\rho }_{\  \mu \nu }$. Classically, in the absence of torsion, $\mathcal H$ satisfies the Bianchi identity
\begin{equation}
\label{E5}
\partial_{[\mu } \mathcal{H}_{\nu \rho \sigma ]}=0.
\end{equation}
In superstring theory, anomaly cancellation through the Green-Schwarz mechanism~\cite{R6}, requires the modified Bianchi identity
\begin{equation}
\label{E6}
\epsilon^{\mu \nu \rho \sigma } \mathcal{H}_{[\nu \rho \sigma ;\mu ]}=\frac{\alpha^{\prime } }{32\kappa } \sqrt{-g} \left( R_{\mu \nu \rho \sigma }\tilde{R}^{\mu \nu \rho \sigma } -F^{a}_{\mu \nu }\tilde{F}^{a\mu \nu } \right)  \equiv \sqrt{-g} \mathcal{G}\left( \omega ,{\bf{A}}\right) 
\end{equation}
where ${\bf{A}}^a$ is a Yang-Mills gauge field with a Latin group index $a$.  Moreover, the gravitationally covariant Levi-Civita symbol is given by
\begin{equation}
\label{E7}
\epsilon^{\mu \nu \rho \sigma } =\frac{\text{sgn} \left( g\right)  }{\sqrt{-g} } \eta^{\mu \nu \rho \sigma } 
\end{equation}
and $\eta^{\mu \nu \rho \sigma } $ is the \emph{flat space} Levi-Civita symbol with $\eta^{0 1 2 3 } =1$. The symbol $\widetilde{\left( \ldots \right)}$ over the curvature or Yang-Mills field strength denotes the tensor dual:
\begin{equation}
\label{E8}
\tilde{R}_{\mu \nu \rho \sigma } =\frac{1}{2} \epsilon_{\mu \nu \lambda \pi } R^{\lambda \pi }_{\  \  \  \rho \sigma },\  \  \tilde{F^{a}}_{\mu \nu } =\frac{1}{2} \epsilon_{\mu \nu \lambda \pi } F^{a\lambda \pi }.\  \ \end{equation} 
In the Euclidean path integral~\footnote{The Levi-Civita symbol $\eta^{\left( E\right)  }_{\mu \nu \rho \lambda } \eta^{\mu \nu \rho \sigma \left( E\right)  } =6\delta^{\sigma }_{\lambda } $; in Minkowski space $\eta_{\mu \nu \rho \lambda } \eta^{\mu \nu \rho \sigma   } =-6\delta^{\sigma }_{\lambda } $. } 
\begin{equation}
\label{ }
Z_{B}=\int D\mathcal{H}\  \exp \left( -S_{B}\right)  
\end{equation}
for the action $S_B$, where the graviton contribution is treated as a background, we incorporate the Bianchi identity~\eqref{E6} through a delta function

\[\prod_{x} \delta \left(\eta^{\mu \nu \rho \sigma } \mathcal{H}_{[\nu \rho \sigma ;\mu ]}\left(x \right) -\mathcal{G}\left( \omega ,{\bf{A}}\right)\right) \]
which can be expressed as a path integral over a pseudoscalar Lagrange-multiplier field, which eventually corresponds to the gravitational (or KR), string-model independent~\cite{kaloper,svrcek}, axion: 
\begin{equation}
\label{E9}
\int Db\exp \left[ i\int d^{4}x\sqrt{-g} \frac{1}{\sqrt{3} } b\left( x\right)   \left(\eta^{\mu \nu \rho \sigma } {\mathcal{H}}_{[\nu \rho \sigma ;\mu ]}\left(x \right) -\mathcal{G}\left( \omega ,{\bf{A}}\right)\right)\right]. \end{equation}
On integrating by parts and on assuming that $\mathcal{H}$ falls off at infinity, the delta function constraint can be re-expressed with the integral 
\begin{equation}
\label{E10B}
\int Db\exp \left[ -i\int d^{4}x\sqrt{-g(x)} \left( \frac{1}{\sqrt{3} } \partial^{\mu } b(x)\eta_{\mu \nu \rho \sigma } H^{\nu \rho \sigma }(x)+\frac{b}{\sqrt{3} }\mathcal{G}\left( \omega ,{\bf{A}}\right) \right)  \right]  
\end{equation}
Hence, on integrating over $\mathcal{H}$, $Z_B$ is
\begin{equation}
\label{E11}
Z_{B}=\int db\exp \left( -\int d^{4}x\sqrt{g^{\left( E\right)  }} \left\{ \frac{1}{2\kappa^{2} } R+\frac{1}{12} \eta^{\left( E\right)  }_{\mu \nu \rho \lambda } \eta^{\mu \nu \rho \sigma \left( E\right)  } \partial^{\lambda } b\partial_{\sigma } b + \, \frac{b}{\sqrt{3} }\mathcal{G}\left( \omega ,{\bf{A}}\right)\right\} \right).  
\end{equation}
We have emphasised the Euclidean formulation by using the superscript $(E)$.The continuation of $Z_{B}$ back to Minkowski space has an ambiguity~\cite{R7}. 
As first stressed in \cite{Mav2020}, 
where microscopic arguments for the emergence of \cPT-symmetric effective theories from strings was given, 
 we have two choices: 
\begin{enumerate}
  \item Before continuing back to Minkowski space  we can replace $\eta^{\left( E\right)  }_{\mu \nu \rho \lambda }\eta^{\mu \nu \rho \sigma \left( E\right)  } $  with $6\delta^{\sigma }_{\lambda } $
  \item After continuing back to Minklowski space we can replace $\eta^{\left( E\right)  }_{\mu \nu \rho \lambda }\eta^{\mu \nu \rho \sigma \left( E\right)  } $  with $-6\delta^{\sigma }_{\lambda }(=\eta_{\mu \nu \rho \lambda } \eta^{\mu \nu \rho \sigma   }) $ and also redefine the phase of $b$ by $\pi/2$ in order to get the canonical sign for the kinetic term. We note that in this later case, the redefinition $b \to i\,b$ is consistent with the postulated transformation of $i\,b$ \eqref{E1d} under time reversal $\mathcal T$, given that the latter transformation would imply $i \to -i$ (the reader is reminded that the initial (hermitian) field $b$ would change sign under $\mathcal T$, $b \to -b$~\cite{RR5}).
  
\end{enumerate}
This ambiguity leads in turn to an ambiguity in the phase of the coefficient of the Chern-Simons anomaly terms in the effective actions appearing in the analytic continuation back to Minkowski spacetime of the exponent of \eqref{E11}.

When fermions are introduced into the model this ambiguity leads to an ambiguity in the phase of the 
derivative of the fermions to the axial current, 
\begin{align}\label{bF}
\mathcal S_{\rm b-F} = {\rm const}\, \times \, \int d^4 x\, \sqrt{-g}\, i^\xi  \, b(x) \, \nabla_\mu \Big( \overline \psi \, \gamma^5 \, \gamma^\mu \, \psi \Big)~\,,
\end{align}
with $\xi=0 ~\rm or~ 1$, depending on the way we analytically continue back. Above, 
$\nabla_\mu$ denotes the usual gravitational covariant derivative, and the constant, in the string-effective action, is determined  in terms of the four-dimensional gravitational constant and the string mass scale~\cite{R6,kaloper}.
If the fermions are chiral, the one-loop anomaly  will lead to a coupling of the axion with Chern-Simons anomaly terms, of a form similar 
to the one above in \eqref{E11}, due to the Green-Schwarz anomaly cancelation mechanism ({\it cf.} Eq.~\eqref{E6}). On the other hand, 
if there are non-chiral fermions present, with mass $m$, say,  then, the classical fermion equations of motion stemming from \eqref{bF} plus the fermion kinetic terms, will lead to a non-derivative Yukawa coupling of the $b$-axion with the pseudoscalar (non-chiral) fermion bilinear, of the type appearing in the Yukawa interactions of \eqref{E1},
\begin{align}\label{bFYuk}
\mathcal S_{\rm b-F-onshell-non-chiral} = 2 \times {\rm const}\, \times  \, \int d^4 x\, \sqrt{-g}\,   \, b(x) \, m\,  \overline \psi \, \gamma^5 \, \psi  =
Y_b \, b(x) \, m\,  \overline \psi \, \gamma^5  \, \psi 
~\,,\end{align}
where the Yukawa coupling can be real or purely imaginary, depending on the above ambiguity on the analytic continuation back to Minkowski spacetime.

\begin{figure}[t]
 \centering
  \includegraphics[width=0.7\textwidth]{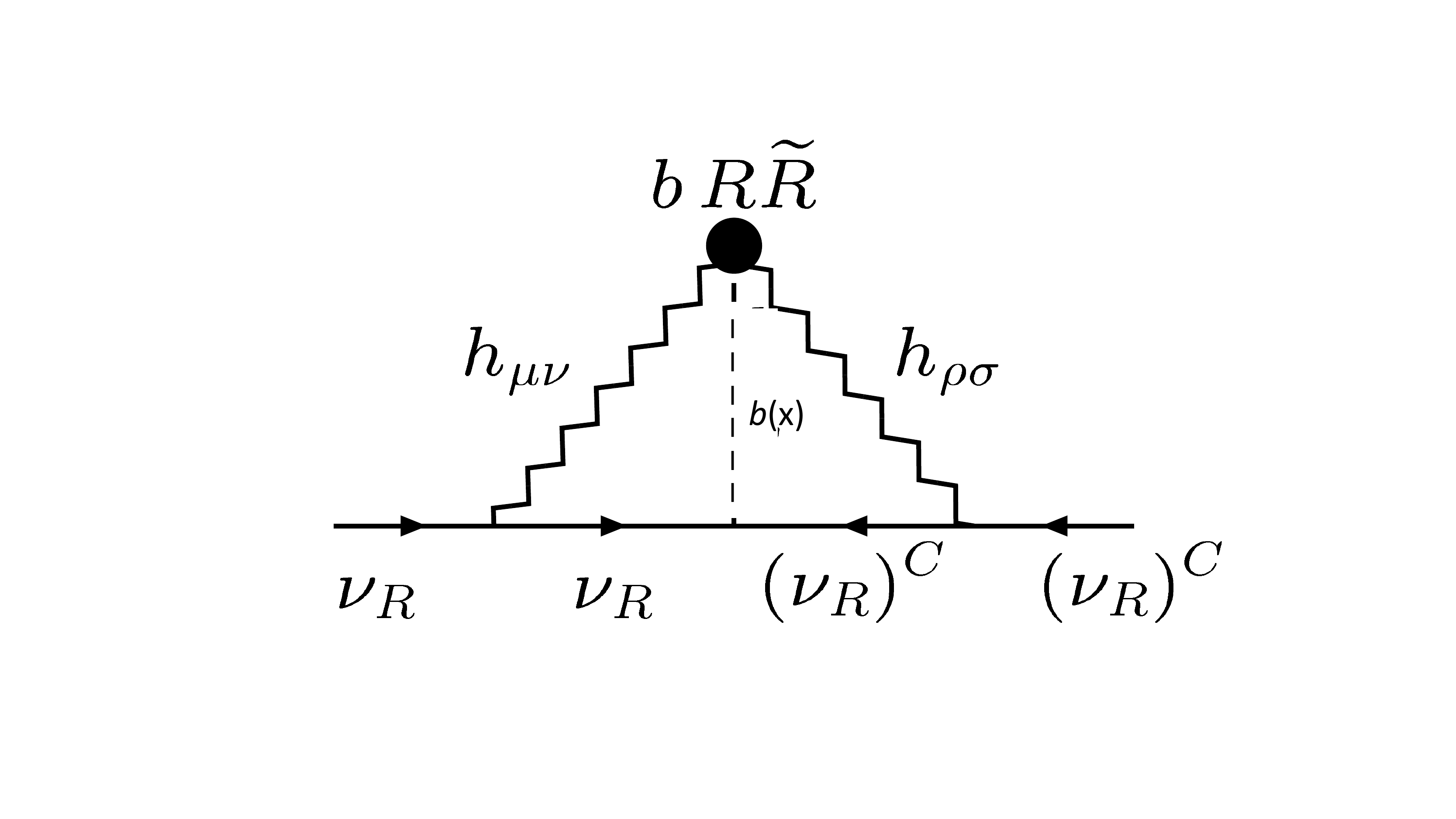} 
  \vspace{-1.5cm}
\caption{\it 2-loop Feynman graph pertaining to the anomalously generated 
 (Majorana) mass for right-handed fermions $\psi_R$, which in the model of \cite{pilaftsis,Mav2020} are identified with sterile neutrinos $\nu_R$.  $C$ denotes the conventional (Dirac) charge conjugation operator. The dark blob denotes the gravitational-anomaly Chern-Simons operator $b(x)\, R_{\mu\nu\lambda\rho}\widetilde{R}^{\mu\nu\lambda\rho}$ (up to numerical coefficients), which, notably, is the only part of the anomaly relevant for sterile neutrinos, since the latter do not couple to gauge fields; 
 $b(x)$ denotes the KR (gravitational) axions. The wavy lines are gravitons $h_{\mu\nu}$, continuous lines with arrows denote the chiral fermions, whilst the dashed lines are the $b$ axions.}
\label{fig:rad}
\end{figure}

On a more technical note, we note~\cite{Mav2020} that, as discussed in \cite{pilaftsis}, it is possible that Yukawa couplings of the form 
\eqref{bFYuk} can be generated by non-perturbative instanton effects, even for \emph{massless} chiral fermions, which can in turn lead, via the mechanism~\cite{pilaftsis} associated with the two-loop diagram of fig.~\ref{fig:rad}, to anomalously-generated radiative masses for the 
chiral fermions, of the form:
\begin{align}\label{chiralmass}
M_F^{\rm chiral} \propto i^\xi \, Y_b\, \kappa^5 \, \Lambda^6 \,, 
\end{align}
where $\Lambda$ is the ultraviolet cutoff of the effective theory (the reader is referred to \cite{pilaftsis} for the omitted real numerical constants of proportionality). 

Thus, we observe~\cite{Mav2020} that for purely imaginary couplings of the $b$-axion with the Chern-Simons terms, 
one may obtain {\it real} dynamical chiral mass generation for {\it real} $i^\xi\, Y_b$, that is 
for purely imaginary Yukawa couplings $Y_b$ \eqref{bFYuk} when $\xi=1$ 
({\it i.e.} non-Hermitian axion-Chern-Simons couplings). 

The models that arise due to this ambiguity are thus either Hermitian or non-Hermitian, of the type \eqref{E1} discussed in~\cite{R4, R5a, R6a, R10}. Such models will be the focus of our discussion in this work but with the important addition of a quartic self-interaction coupling for the pseudoscalar field, which was not considered in \cite{R4, R5a, R6a, R10}, but included in \cite{R4}. The role of self-interactions in dynamical generation of fermion and pseudoscalar masses will be examined using a nonperturbative analysis.

\section{The sign of non-Hermiticity}
\label{e1b}
The Lagrangian $\mathcal {L}$ in \eqref{E1} represents a renormalisable Lagrangian in 4-dimensions involving pseudoscalars and fermions. The relation of renormalisation to the emergence of \cPT symmetry is an issue which has been noted previously~\cite{RR14,RR14a, RR15}. In early studies of dispersion relations in quantum field theory it was shown that a formally Hermitian theory may contain ghosts~\cite{RR16}.~\cPT symmetry, if present, may allow interpretations where ghosts (appearing in conventional interpretations) are absent. A~\cPT-symmetric interpretation restored stability to the vacuum for a conventionally unstable Higgs vacuum field in the SM~\cite{RR17}.\footnote{In that work~\cite{RR17} the effective potential $\Gamma \left( \varphi_{c} \right) $  was treated as a function of the one dimensional variable $\varphi_{c}$ and the potential was studied using the techniques of quantum mechanics. The effective potential, which arose out of an evaluation of a fermion functional determinant~\cite{R13}, has the form:
\begin{equation}
\label{e1c}
\Gamma \left( \varphi_{c} \right) \propto -\varphi^{4} \rm ln\varphi^{2} 
\end{equation} 
for large $\varphi_{c}$. It is an  upside down potential that is a  \cPT symmetric potential which has arisen form a Hermitian theory due to quantum corrections.}

Starting with Hermitian couplings we will review in this section evidence from several  (perturbative) renormalisation group analyses~\cite{R4,RR11,RR12,RR13} which indicate that there may be flows towards \cPT-symmetric fixed points. With the discovery of non-Hermitian but \cPT-symmetric unitary quantum mechanics~\cite{R2}, such flows may indicate one way that \cPT symmetric field theories are required. 
Below we review the main points of this analysis, first at one loop order, and then we extend the discussion to the two-loop case, where in the massless theory we demonstrate a renormalisation group (RG)  flow between Hermitian and non-Hermitian fixed points.


\subsection{The one loop renormalisation group flow for $\lambda$ and $g$\label{sec:group}}

As shown in~\cite{R4}, the RG equations~\cite{RR18} associated with (the dimensionally regularised)  $\mathcal{L}$ in $D=4-\epsilon$ dimensions,
\begin{eqnarray}
\frac{dg}{dt} & = &\beta_{g} \left( g\right) = \frac{5g^{3}}{16\pi^{2}}-\frac{\epsilon g}{2} \label{e11} \\
\frac{d\lambda}{dt} & =&\beta_{\lambda} \left( g, \lambda\right)= \frac{48 g^{4}-3\lambda^{2}+8g^{2}\lambda}{16\pi^{2}} -\lambda\epsilon\label{e12}\\
\frac{dm}{dt} & = &\beta_{m} \left( g, m\right)=-\frac{g^{2}m}{16\pi^{2}}\label{e13}\\
\frac{dM}{dt} & = &\beta_{m} \left( g, m, M\right)=\frac{1}{32\pi^{2}M}\left[4g^{2} (M^{2}-2m^{2}) -\lambda M^{2}   \right] \label{e14}
\end{eqnarray}
where we have given explicit expression for the renormalisation group $\beta$ functions and  $\frac{d}{dt} =\mu \frac{d}{d\mu } $, $\mu$ being the mass scale  used in the method of dimensional regularisation~\cite{RR18,RR19}.

 In the present work, where our focus is on dynamically generated masses, we restrict our attention to models with zero bare masses, and hence the last two RG equations in \eqref{e14} are trivial and so are ignored.
Moreover the first two one-loop RG equations decouple from the rest, and thus the fixed-point structure of the interaction couplings $(g, \lambda)$ can be determined by concentrating on these two equations. Below we repeat the main conclusions of \cite{R4} in this respect. For more details we refer the interested readers to that work.

The fixed points of $g$ are $g^*$  where  $g^*=g^*_{\pm}=\pm \sqrt{\frac{8 \pi^{2}\epsilon}{5}}$ and the trivial fixed point $g^*=0$.\footnote{The sign of $g$ distinguishes separate parts of "theory" space. } The related fixed points $\lambda^*$ for $\lambda$ are determined by
\begin{equation}
48{g^{*}}^{4}-3{\lambda^{*}}^{2}+8{{g^{*}}^{2}}\lambda^{*}=16\pi^{2}\epsilon {\lambda^{*}}.
\label{e17}
\end{equation} 

The solutions for $\lambda^{*}$  are $\lambda^{*}=0$ and $\lambda^{*}_{\pm}=\lambda_{\pm}\epsilon$ where

\begin{equation}
\label{e18}
\lambda_{\pm}=\frac{8}{3}g_{0}^{2} \pm \sqrt{\frac{64}{9}g_{0}^{4}+64}
\end{equation}
and $g_{0}=\sqrt{\frac{8\pi^{2}}{5}}\sim 3.97$, which gives $\lambda_{+}\sim 84.97$ and $\lambda_{-}\sim -0.75$. Thus, we 
observe that $\lambda_{-}$ is negative and, therefore 
 is a Hermitian fixed point. On the other hand,  $\lambda_{+}$ is positive and, thus, is a non-Hermitian fixed point.\footnote{ We note  that in the context of related Hermitian models involving massless fermions with Yukawa interactions and a Higgs sector with scalar self-interactions,
 the possibility of a flow to a non-Hermitian quartic self-coupling fixed point has also  been noticed~\cite{RR13}.}
 At the level of fixed points, non-Hermiticity is therefore introduced through the $\lambda$ coupling. On the other hand, $g$ remains real at the fixed points. 

 As discussed in \cite{R4}, and reviewed below,  the pertinent $\epsilon$-dependent fixed points are examples of Wilson-Fisher fixed points \cite{RR19}, and their linear stability has been examined in that work, 
to which we refer the interested reader for details.

We shall consider the solutions of the coupled flow equations  \eq{e11} and \eq{e12} (and the closely related equations \eq{e17} and \eq{e18}). We can rewrite \eq{e11} as
\begin{equation}
\label{e20}
\frac{dg}{dt}=\frac{5g}{16\pi^{2}}(g-g_{+})(g-g_{-}).
\end{equation}

For $g\gg g_{+}$ \eq{e20} simplifies to
\begin{equation}
\label{e21}
\frac{dg}{dt}=\frac{5}{16\pi^{2}}g^3
\end{equation}
and leads to 
\begin{equation}
\label{e22}
y\equiv g^2=-\frac{1}{2(c+\frac{5t}{16\pi^{2}})}
\end{equation}
where  $c$ is a constant of integration. At $t=0$, if the theory is Hermitian, then $c$ is negative. As $t$ increases, $g$ increases  but remains Hermitian until at finite time $t=\frac{16\pi^{2}|c|}{5}$ the approximation of small $g$, and thus perturbative renormalisation breaks down. 

For $g\ll{g_{-}}$ we again  have \eq{e21} and $c$ is negative for a theory which is Hermitian at a scale $\mu \sim 1$. In the IR, $g$ remains small. In the UV, $g$ moves towards $g=0$ but then veers  away to large positive values of $g$ where perturbation theory is not trustworthy.

For $0<g<g_{+}$ it is clear that $g  \to 0$ as $t \to \infty$. As $t \to -\infty$ we have $g \to g_{+}$. As $\epsilon \to 0$ there is a bifurcation where the fixed points $g_{+}, g_{-}, 0$ coalesce. The trivial fixed point is unstable both in the IR and the UV.

We will now consider the flow of $\lambda$ using \eq{e12}. The solution of \eq{e20} is 
\begin{equation}
\label{e33}
y(t)=-\frac{8d(\epsilon)\pi^{2}\epsilon}{5(e^{\epsilon t}-d(\epsilon)}
\end{equation}
where $d(\epsilon)=e^{8\pi^{2}\epsilon c_{1}}(>0)$ and $c_{1}$ is a constant of integration. The resultant solution of \eq{e12} for $\lambda(t)$ is 
\begin{equation}
\label{e34}
\lambda(t)=8 d(\epsilon){\pi^{2}}\epsilon (1+\sqrt{145}-\frac{c_{2}(\sqrt{145}-1)e^{\sqrt{29/5}\epsilon t}}{(e^{\epsilon t}-d(\epsilon))^{\sqrt{29/5}})})/({15(e^{\epsilon t}-d(\epsilon)) (1+\frac{c_{2}e^{\sqrt{\frac{29}{5}}\epsilon t}}{(e^{\epsilon t}-d(\epsilon))^{\sqrt{29/5}}})})
\end{equation}
where $c_2$ is an integration constant.
This is complicated to analyse. If we keep away from the region of the fixed points near the origin, by considering  $\epsilon \to 0$, the solutions in \eq{e33} and \eq{e34} can be simplified to
\begin{equation}
\label{e35}
y(t)=-\frac{1}{\frac{5}{8\pi^{2}} t+c_{1}}
\end{equation}
and
\begin{equation}
\label{e36}
\lambda(t)=\frac{8\pi^{2}\left[1-\sqrt{145}+\left(1+\sqrt{145}\right)\left(8c_{1}\pi^{2}+5t\right)^{\sqrt{\frac{29}{5}}}c_{2}\right]}{3(8c_{1}\pi^{2}+5t)(1+c_{2}(8c_{1}\pi^{2}+5t)^{\sqrt{29/5}})}.
 \end{equation}
For $g$ to be non-Hermitian at $t=0$, we have $y<0$, and so $c_{1}>0$; so as $t \to \infty$, $y$ remains non-Hermitian but slowly vanishes. In the infrared (IR), as $t$ decreases from $t=0$, $y$ increases but remains non-Hermitian; perturbation theory becomes unreliable.

In the Hermitian case, $y>0$ at $t=0$ and so $c_1$ is negative. As $t$ increases from $t=0$ $y$ remains Hermitian but increases until perturbation theory is invalid.

For $\lambda$ to be real we need $(8 c_{1}\pi^{2}+5 t)$ to be nonnegative. This requires $c_2=0$ and so 
\begin{equation}
\label{e37}
\lambda(t)=-\frac{8\pi^{2}\left(\sqrt{145}-1\right)}{3\left(8c_{1}\pi^{2}+5t\right)}.
 \end{equation}
 
 The implication of a non-Hermitian $g$ ($c_{1}>0$) for $\lambda$ is that it is Hermitian ( i.e. $\lambda <0$ ) at $t=0$. As $t \to \infty$ , $\lambda$ falls-off to $0$ but remains Hermitian. In the IR, $\lambda$ increases until perturbation theory is unreliable.  
 
  The implication of a Hermitian $g$ ($c_{1}<0$) is that $\lambda$ is non-Hermitian ( i.e. $\lambda>0$ ) and remains so in the IR. The self-interaction coupling $\lambda$ increases in the UV  until the perturbative analysis becomes unreliable. In the IR, $\lambda$ falls-off but remains non-Hermitian.\footnote{The beta functions for the $g$ and $u$ parameters do also decouple from those for the couplings $m$ and $M$ at two loops.}

\subsection{Two-loop renormalisation group analysis for the massless theory: Renormalisation Group Flows between Hermitian and non-Hermitian fixed points\label{sec:2loop}}

The presence of both Hermitian and non-Hermitian fixed points within our models could  be the result of the one loop nature of our approximation. It is of  course, in general, difficult to rule out this possibility without some parameter in the theory which can control the contributions of higher loops.  However we have analysed a two loop renormalisation flow~\cite{RR11,RR12} for a similar, but massless, Yukawa model given by the Lagrangian $\mathcal{L}_{MY}$\footnote{It  is of course possible to do the analysis with nonzero $m$ and $M$ and it straightforward to show that
\begin{equation*}
\frac{dm}{dt} =\frac{11g^{4}m}{1024\pi^{4} } -\frac{g^{2}m}{16\pi^{2} } 
\end{equation*}
and
\begin{equation*}
\frac{dM^{2}}{dt} =\frac{-8g^{2}m^{2}+(4g^{2}-u)M^{2}}{16\pi^{2} } -\left(\frac{5u^{2}M^{2}}{6} -32g^{4}m^{2}+2g^{4}M^{2}-4g^{2}u\left( m^{2}+M^{2}\right)\right)/256 \pi^{4} . 
\end{equation*}
So $m=M=0$ is a solution and we explore the resultant massless theory, which is the case relevant for the study of dynamical mass generation. } In what follows, we shall demonstrate that, in such a model, there is a renormalisation-group flow between Hermitian and non-Hermitian fixed points. 

 The Lagrangian $\mathcal{L}_{MY}$ is:

\begin{equation}
\label{e44a }
\:\mathcal{L}_{MY}=\frac{1}{2}\left(\partial\phi\right)^{2}+i\bar{\psi}\gamma^{\mu}\partial_{\mu}\psi-i\,g\,\phi\bar{\psi}\gamma_{5}\psi -\frac{u}{4!}\phi^{4}
\end{equation}
where, $\psi$ is a massless Dirac-fermion field and $\phi$ is a massless pseudoscalar field, $g$ denotes the Yukawa coupling,  and $u$ denotes the self-interaction of $\phi$. We shall consider $u>0$ (the Hermitian case) but allow $g$ to be real or imaginary. From the consideration of the convergence of path integrals given earlier we know that the usual Feynman rules are valid. If $u$ were to go towards a negative $u$ fixed point, according to the renormalisation group flow, then for a resulting $\left| u\right|  $ which is not small might be indicative of an emergence of non-Hermiticity. If $\left| u\right|  $ is small then the Feynman rules would be still valid  since the Feynman rules give an approximation to the behaviour near the trivial saddle point of the path integral.

We define for notational convenience 
\begin{align}\label{newdefgh}
\tilde g \equiv \frac{g^2}{16\pi^2} \quad  {\rm and} \quad   
h \equiv \frac{u}{16\pi^2}~.
\end{align}
The loop calculation involves 14 topologically distinct graphs. In \cite{RR11,RR12} the calculation of the beta function $\beta_{\tilde g}$ for $\tilde g$ gives~\footnote{In $D=4-\epsilon$ the beta functions for the couplings in the model would have $\epsilon$ dependent terms determined by the engineering dimensions of the couplings in the noninteger $D$ dimension.} 
\begin{equation}
\label{e45a}
\beta_{\tilde g}=10\, \tilde g^{2}+\frac{1}{6}\, h^{2} \, \tilde g - 4\, h\, \tilde g^{2}-\frac{57}{2}\, \tilde g^{3}
\end{equation}
and the calculation of the  beta function $\beta_{h}$ for $h$ gives
\begin{equation}
\label{e46a}
\beta_{h}=3\, h^{2} + 8\, h\, \tilde g - 48\, \tilde g^{2}-\frac{17}{3}\,h^{3}-12\, \tilde g\, h^{2}+28\, h\, \tilde g^{2}+384\, \tilde g^{3}.
\end{equation}

We can show that there are four fixed points $(\tilde g_{i},h_{i}), i=1,\cdots,4$ where
\begin{eqnarray}
\tilde g_{1} & = & 0 \, \quad \qquad \qquad \qquad h_{1} = 0\label{e51a}  \\
\tilde g_{2} & = &0 \, \quad \qquad \qquad \qquad h_{2} = 0.529412\label{e51b}   \\
\tilde g_{3} & = & -0.00570795 \, \qquad h_{3} = 0.525424\label{e51c}   \\
 \tilde g_{4}& = & 0.234024 \qquad  \qquad h_{4} =  1.01657 \label{e52}
\end{eqnarray}

In this two loop  calculation we note the appearance of a non-Hermitian (purely imaginary ({\it cf.} \eqref{newdefgh}) Yukawa coupling $g$ at  the $i=3$ fixed point.

The possible connection between Hermitian and non-Hermitian fixed points that we have noticed is unlikely to be an artefact. There is some independent evidence that this happens in other theories although the possible connection with \cPT symmetry  was not realised. This independent evidence has been found  in a more complicated model, a chiral Yukawa model~\cite{RR13}, with the Standard Model symmetry implemented only at the global level.The flow of the quartic scalar coupling from positive to negative values was observed. Furthermore the existence of infrared fixed points has a  bearing on a nonperturbative treatment of dynamical symmetry breaking,

\section{Schwinger-Dyson equations}\label{sec:SD}

We have argued that \cPT symmetry can not only be built into model building but can also arise from renormalisation. 

We will now consider dynamical mass generation in \cPT-symmetric field theories. and point out new features not present in Hermitian field theories. We are going to discuss the cases of  weak Yukawa and self-interaction couplings, near the trivial fixed point \eqref{e51a}. It should be stressed that the conclusions derived here, will be modified once we perform the analysis in the neighborhood of the other non-trivial fixed points, which is the topic of another work.

 The SD equations can be derived in terms of the (Euclidean)  path integral  (for the partition function) $Z [J, \eta, \bar{\eta}]$ :

 \begin{eqnarray}
Z [J, \eta, \bar{\eta}] &=& \int \mathcal{D} [\phi \psi \bar{\psi}] \exp
  \left\{ -\int d^4 x \left[ \frac{1}{2} \partial_{\mu} \phi \partial^{\mu}
  \phi +\frac{M^2}{2} \phi^2+ \bar{\psi} i \slashed{\partial} \psi + i g \phi \bar{\psi} \gamma^5 \psi
  - \frac{\lambda}{4!} \phi^4 \right] \right.\nonumber\\
& & \left.+  \int d^4 x [J \phi + \bar{\psi} \eta +
  \bar{\eta} \psi] \right\}\label{e53}
\end{eqnarray}
where $\psi$ and $\bar{\psi}$ are Grassman field variables, $\eta$  and $\bar{\eta}$ are Grassman sources, and~$\phi$ and $J$  are c-number field and source  respectively.

The scalar SD equations are derived from the functional equation:
\begin{equation}
\label{e54}
  \int \mathcal{D} [\phi \psi \bar{\psi}] \frac{\delta}{\delta \phi (x)} e^{-
  S} = 0.
\end{equation}
The fermion SD equations are derived from the functional equation:
\begin{equation}
\label{e55}
  \int \mathcal{D} [\phi \psi \bar{\psi}] \frac{\delta}{\delta \bar{\psi}}
  e^{-S} = 0.
\end{equation}
\subsection{\cPT symmetry and the scalar SD equation}

Based on conventional perturbation theory, quantisation of $\mathcal L$ does not lead to odd-point Greens functions. Such odd-point Green's functions typically arise in \cPT symmetric scalar field theories~\cite{RR20,R2}. For completeness we consider whether our conventional Schwinger-Dyson analysis is affected by these odd-point Green's functions. We illustrate this for the field theory with $g=0$ where the path integral reduces to that for a scalar \cPT symmetric field  theory. The connected $n$-point Green's functions in the presence of a current $J\left(x\right)$ are defined by
\begin{equation}
\label{e56}
G^{\left( J\right)  }_{n}\left( x_{1},x_{2},\ldots ,x_{n}\right)  \equiv \frac{\delta^{n} }{\delta J\left( x_{1}\right)  \delta J\left( x_{2}\right)  \ldots \delta J\left( x_{n}\right)  } \ln\left( Z\left[ J\right]  \right)  
\end{equation}
and $Z\left[ J\right]$ is the vacuum persistence amplitude $\langle 0|0\rangle_{J}$ where $|0\rangle$ is the vacuum ket. Let $\tilde{\lambda } \equiv \frac{\lambda}{6}$. From \eqref{e54} we deduce that
\begin{equation}
\label{e57}
-\partial^{2} \phi \left( x\right)  +M^{2}\phi \left( x\right)  -\tilde{\lambda } \phi \left( x\right)^{3}  =J\left( x\right).  
\end{equation}
The one-point Green's function in the presence of the source is defined as 
\begin{equation}
\label{e58}
G^{\left( J\right)  }_{1}\left( x\right)  \equiv \frac{\langle 0|\phi \left( x\right)  |0\rangle_{J} }{Z\left[ J\right]  } .
\end{equation}
On taking the vacuum expectation value of the terms in \eqref{e57} we have
\begin{equation}
\label{e59}
-\partial^{2} G^{\left( J\right)  }_{1}\left( x\right)  +M^{2}G^{\left( J\right)  }_{1}\left( x\right)  -\tilde{\lambda } \frac{\langle 0|\phi^{3} \left( x\right)  |0\rangle_{J} }{Z\left[ J\right]  } =J\left( x\right).  
\end{equation}
We can rewrite \eqref{e58} as
\begin{equation}
\label{e58a}
G^{\left( J\right)  }_{1}\left( x\right) Z\left[ J\right] =\langle 0|\phi \left( x\right)  |0\rangle_{J}
\end{equation}
and take functional derivatives with respect to $J\left( x\right)$. It is straightforward to show that 
\begin{equation}
\label{e60}
\left[ G^{\left( J\right)  }_{1}\left( x\right)  \right]^{3}  Z\left[ J\right]  +3G^{\left( J\right)  }_{1}\left( x\right)  G^{\left( J\right)  }_{2}\left( x,x\right)  Z\left[ J\right]  +G^{\left( J\right)  }_{3}\left( x,x,x\right)  Z\left[ J\right]  =\langle 0|\phi^{{}^{3}} \left( x\right)  |0\rangle_{J}, 
\end{equation}
a result which is used in  \eqref{e59} to give
\begin{equation}
\label{e61}
-\partial^{2} G^{\left( J\right)  }_{1}\left( x\right)  +M^{2}G^{\left( J\right)  }_{1}\left( x\right)  -\tilde{\lambda } \left( \left[ G^{\left( J\right)  }_{1}\left( x\right)  \right]^{3}  +3G^{\left( J\right)  }_{1}\left( x\right)  G^{\left( J\right)  }_{2}\left( x,x\right)  +G^{\left( J\right)  }_{3}\left( x,x,x\right)  \right)    =J\left( x\right).  
\end{equation}

From this equation the other Schwinger-Dyson equations is obtained by functional differentiation with respect to $J\left( x\right)$ and then setting $J\left( x\right)=0$. There is an infinite chain of equations which means that it is necessary to truncate the chain by making an assumption that $G^{\left( J\right)  }_{n}\left( x\right)=0$ for $n>n_{0}$ for some positive integer $n_{0}$. It is known that~\cite{RR20} for $n_{0}=2$
\begin{equation}
\label{e62}
-\partial^{2} G^{\left( J\right)  }_{1}\left( x\right)  +M^{2}G^{\left( J\right)  }_{1}\left( x\right)  -\tilde{\lambda } G^{\left( J\right)  }_{1}\left( x\right)^{3}  w\left( \gamma^{\left( J\right)  } \right)  =J\left( x\right)
\end{equation}
where 
\begin{equation}
\label{e63}
w\left( y\right)  \equiv \Gamma \left( 5\right)  \sum^{2}_{k=0} \frac{\left( -1\right)^{k}  }{\Gamma \left( 5-2k\right)  2^{k}\Gamma \left( k+1\right)  y^{2k}} 
\end{equation}
and 
\begin{equation}
\label{e64}
\gamma^{\left( J\right)  } \left( x\right)  =\frac{iG^{\left( J\right)  }_{1}\left( x\right)  }{\sqrt{G^{\left( J\right)  }_{2}\left( x,x\right)  } } .
\end{equation}

When $J$ is set to $0$, $G^{\left( J\right)  }_{1}\left( x\right)=G_{1}$ and $G_{1}$ is independent of $x$ and $\gamma^{\left( J\right)  }$ is written as the constant  $\gamma_{D}$ (in $D$  dimensions) which is 
\begin{equation}
\label{e64a}
\gamma_{0} =\frac{iG_{1}}{\sqrt{G_{2}\left( 0\right)  } } .
\end{equation}
For nontrivial solutions $\gamma_{0}$ needs to be a zero of $w(x)$.
For $D=0$ and a \emph{massless} theory~\eqref{e62}  is  solved to find~\cite{RR20} a \emph{nontrivial} solution for $G_{1}$:
\begin{equation}
\label{e65}
G_{1}=-i\left( \frac{4}{\tilde{\lambda } } \right)^{1/4}  \frac{\Gamma \left( \frac{3}{4} \right)  }{\sqrt{\pi } } .
\end{equation}
 It is interesting to note that $G_{1}$ is imaginary, a feature actually valid in \emph{any} dimension\footnote{Recently it has been argued using a path integral formulation}. The (truncated) equation for $G_{2}\left( x-y\right)$ is
\begin{equation}
\label{e67}
-\partial^{2} G_{2}\left( x-y\right)  +M^{2}G_{2}\left( x-y\right)  -3\tilde{\lambda } G^{\  2}_{1}\tilde{w} \left( \gamma \right)  G_{2}\left( x-y\right)  =\delta \left( x-y\right)  
\end{equation}
where 
\begin{equation}
\label{e67a}
\tilde{w} \left( y\right)  \equiv 2\sum^{1}_{k=0} \frac{\left( -1\right)^{k}  }{\Gamma \left( 3-2k\right)  2^{k}\Gamma \left( k+1\right)  y^{2k}} .
\end{equation}
and $G_{2}\left( x-y\right)$ is real. The analysis is again simple for $D=0$ and it can be shown that 
\begin{equation}
\label{e68}
G_{2}=\left( \frac{1}{3\tilde{\lambda } \gamma^{2}_{0} \tilde{w} \left( \gamma_{0} \right)  } \right)^{1/2}  
\end{equation}
which is real. For general $D$ we can see that for a massless theory (i.e. one with $M=0$) there is an effective mass $\tilde m$ represented by $-3\tilde{\lambda } G^{\  2}_{1}\tilde{w} \left( \gamma \right)$ which is positive since $G_{1}$ is pure imaginary. If in $D=4$ this mass contribution is exponentially small then a conventional SD analysis will be adequate. We shall see however that in $D=1$ that this mass is not  exponentially small.

 For $D=1$ it can be shown that~\cite{RR20} 
\begin{equation}
\label{e69}
\tilde{m}^{2} =\sqrt{\frac{2\tilde{\lambda }^{2} }{\tilde{m} } -\frac{4\tilde{\lambda }^{4} }{9\tilde{m}^{6} } +\frac{7\tilde{\lambda }^{6} }{24\tilde{m}^{11} } +\ldots } 
\end{equation}
and solving this for $\tilde m$ it can be shown that 
\begin{equation}
\label{e70}
\tilde m\approx 1.126151 \, \tilde \lambda^{2/5}.
\end{equation}
There are two things to notice about this result
\begin{itemize}
  \item the form for $\tilde m$ is noperturbative.
  \item the functional dependence on $\tilde \lambda$ is different from the usual dynamical mass generation where the masses fall off as $\exp \left( -\frac{1}{\tilde{\lambda } } \right) \,.$ 
   \end{itemize}
If this feature were to persist for $D=4$ then our conventional SD  analysis in $D=4$ would need revision. We shall adapt the discussion above  $G_{1}$ in $D=4-\epsilon$ for a massless theory. In \cite{RR20} for  $D=4-\epsilon$, for a truncated theory as discussed above, it was shown that 
\begin{equation}
\label{e71}
G_{1}=-i\left\{ \left[ 3\tilde{\lambda } \tilde{w} \left( \gamma_{0} \right)  \right]^{\frac{2-\epsilon }{2} }  \left[ \Gamma \left( -1+\frac{\epsilon }{2} \right)  \gamma^{4}_{0} \left( 4\pi \right)^{-4+\epsilon }  \right]  \right\}^{\frac{1}{2\epsilon } }. 
 \end{equation}
Although this is not rigorous, the above expression for  $G_{1}$ indicates the  it vanishes faster than any power of $\tilde{\lambda }$ as $D=4$ is approached, i.e. as $\epsilon \to 0+$. Moreover, if any mass~$M$ is generated by  some  mechanism other than the \cPT-symmetric mechanism (PTSM) discussed above, then PTSM
will produce a $G_{1}$ of the order of $\exp \left( -\frac{1}{\tilde{\lambda } } \right)  $. Hence, since our interest is in dynamical mass generation in $D=4$, we shall not consider PTSM and the ensuing odd-point Green's functions in our analysis given in the next section.

\subsection{Schwinger-Dyson equations for the Yukawa theory in $D=4$}
 We derive the standard SD equations for the theory described by  $\mathcal{L}$ from  \eqref{e53} (on ignoring odd-point (pseudo)scalar Green's functions as discussed above).
  Although the derivation is done using the Euclidean formalism, as required for the formal convergence of the path integral, we eventually analytically continue back to Minkowski spacetime. In what follows, we therefore quote (formally) the SD in Minkowski spacetime.The details are given in Appendix \ref{sec:appC}. 
  
We first introduce the functionals $W$ and $\Gamma$ defined by
\begin{equation}
\label{A1}
Z=e^{-i W}
\end{equation}
and then the functional $\Gamma$ (through a Legendre transformation)
\begin{equation}
\label{A1a}
  W [J, \eta, \bar{\eta}] = - \Gamma [\phi \psi \bar{\psi}] - \int d^4 x [J
  \phi + \bar{\psi} \eta + \bar{\eta} \psi].
\end{equation}
The inverse scalar and fermion propagators are
\begin{equation}
\label{A2}
i G^{- 1}_s (y -z)=\frac{\delta^2 \Gamma}{\delta \phi (y) \delta \phi (z)},  
\end{equation}
and
\begin{equation}
\label{A3}
-i G^{- 1}_f (z - y)=\frac{\delta^2 \Gamma}{\delta \bar{\psi} (y) \delta \psi (z)} .
\end{equation}
The proper Yukawa vertex is 
\begin{equation}
\label{A4}
\Gamma^{(3)} (r, v, w)\equiv\frac{i \delta^3 \Gamma}{\delta \phi (r) \delta \bar{\psi} (v) \delta\psi (w)} .
\end{equation}
The proper 4-scalar vertex is 
\begin{equation}
\label{A5}
\Gamma^{(4)} (r, r', v, w)\equiv\frac{i \delta^4 \Gamma}{\delta \phi (r) \delta \phi (r') \delta \phi (v)
\delta \phi (w)}\, .
\end{equation}

\begin{figure}[t]
\centering
\includegraphics[width=0.7\textwidth]{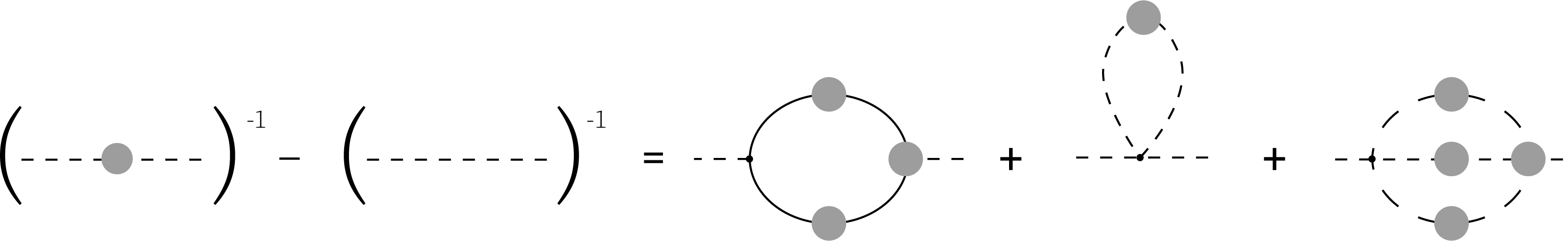} \hfill \includegraphics[width=0.7\textwidth]{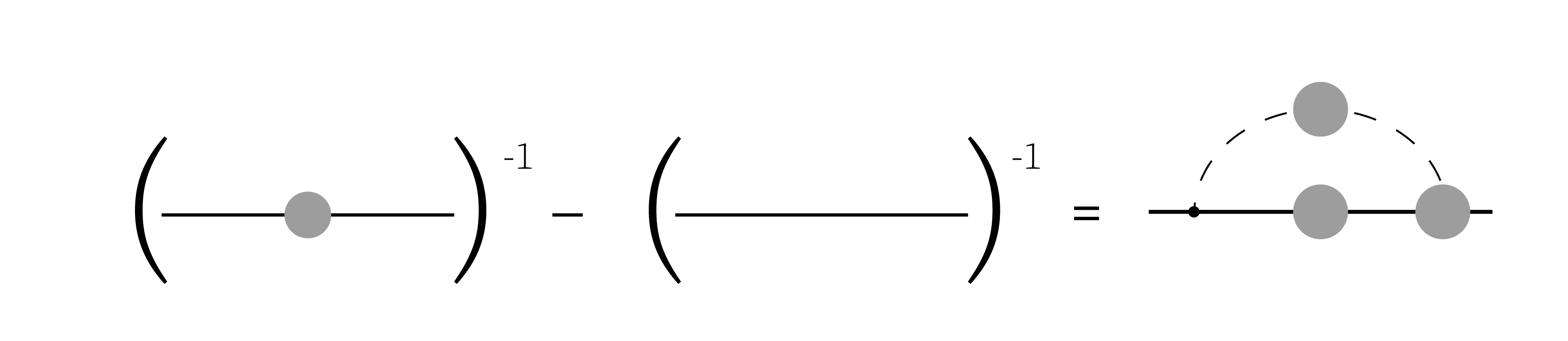} 
\caption{Dressed (inverse) propagators for fermion (upper diagram) and pseudoscalar (lower diagram) fields in the theory \eqref{e53}. Dashed lines are (pseudoscalar) fields, whilst continuous lines are fermions.The schematic grey blobs denote 
quantum corrections.}
\label{fig:fsprops}
\end{figure}

\begin{figure}[t]
\centering
\includegraphics[width=0.7\textwidth]{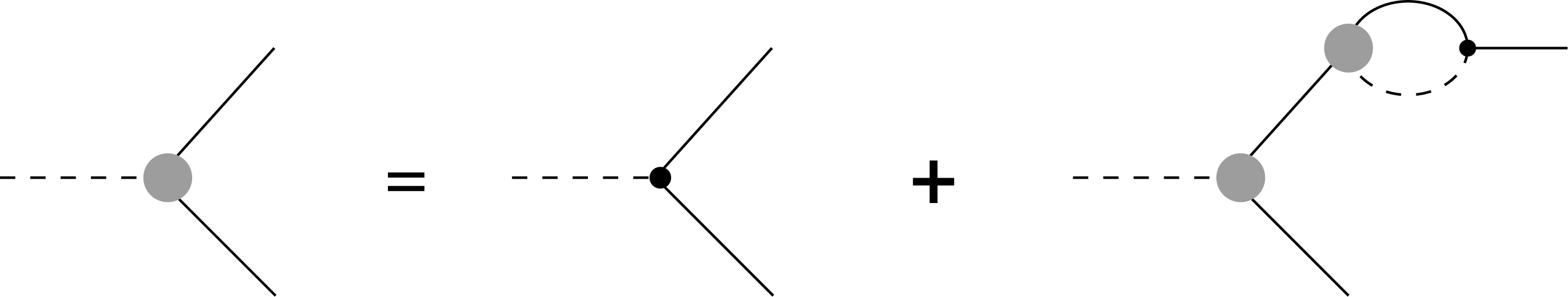} \hfill \includegraphics[width=0.7\textwidth]{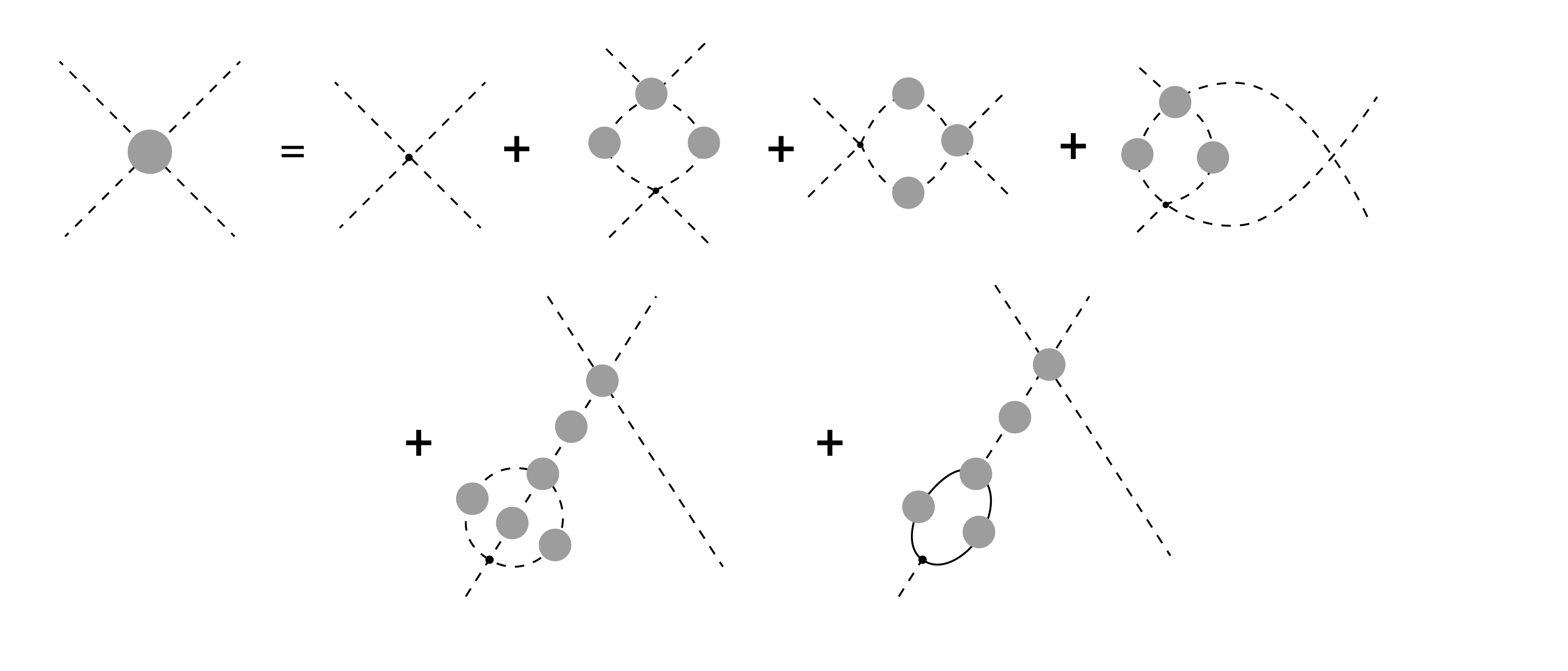} 
\caption{Dressed vertices: $\phi \overline \psi \, \psi $ (upper diagram) and $\phi^4$ (lower diagram) fields in the theory \eqref{e53}. Dashed lines are (pseudoscalar) fields, whilst continuous lines are fermions. The grey blobs denote schematically quantum corrections. The dark point refers to tree-level vertex.}
\label{fig:vert}
\end{figure}

In terms of these quantities, on following standard methods outlined in Appendix \ref{sec:appC}, we find coupled SD equations (with the truncation of ignoring
$n$-point Green's functions for $n \ge 5$). 
In particular, we obtain  the following equations for (pseudo)scalar and fermion propagators, respectively (see fig.~\ref{fig:fsprops}):
\begin{eqnarray}
\label{A7}
  G^{- 1}_s (k) - S^{- 1}_s (k) &=& \tmop{tr} \left[ \int_p g \gamma^5 G_f (p)
  \Gamma^{(3)} (p, k) G_f (p - k) \right] - i \frac{\lambda}{2} \int_p G_s (p) \nonumber\\
& & - i \frac{\lambda}{3!} \int_p \int_l G_s (p) G_s (k + l - p) \Gamma^{(4)} (k, p, l) G_s (l) \,,
\end{eqnarray}
\begin{equation}
\label{A6}
  G^{- 1}_f (k) - S^{- 1}_f (k) = - \int_p g \gamma^5 G_f (p)
  \Gamma^{(3)} (p, k) G_s (p - k) \,,
\end{equation}
where:
\begin{equation}
\label{A8}
\Gamma^{(3)} (q, p, p') = g \gamma^5 - \int_k [g \gamma^5 G_f (k)
\Gamma^{(3)} (k, p) G_s (p - k)] G_f (q - p') \Gamma^{(3)} (q, q - p', p')\,,
\end{equation}
and
\begin{eqnarray}
\label{A9}
\Gamma^{(4)} (q, q', p, p') &=& i \lambda\nonumber\\
& & + i \frac{\lambda}{2} \int_k G_s (q + q' - k) \Gamma^{(4)} (q, q', k, q + q' - k)G_s (k)\nonumber\\
& & + i \frac{\lambda}{2}  \int_k G_s (k) \Gamma^{(4)} (k, q', k + p - q, p') G_s (k +p - q)\nonumber\\
& & + i \frac{\lambda}{2} \int_k G_s (k) \Gamma^{(4)} (q, q' + k - p, k, p') G_s (q' +k - p)\nonumber\\
& & - i \frac{\lambda}{3!} \int_k \int_{k'} G_s (k) G_s (k') \Gamma^{(4)} (k + k' - p, q
+ q' - p', k, k') G_s (k + k' - p) \nonumber\\
& & \times G_s (q + q' - p') \Gamma^{(4)} (q, q', q + q' - p', p')\nonumber\\
& & + \int_k \tmop{tr} [g \gamma^5 G_f (k) \Gamma^{(3)} (k, p) G_f (k - p)] G_s
(q + q' - p') \Gamma^{(4)} (q, q', q + q' - p', p')\,,
\end{eqnarray}
for the vertices $\phi\bar{\psi}\psi$ and $\phi^4$, respectively (see fig.~\ref{fig:vert}).

These equations are nonlinear integral equations which  are regularised with a momentum-cut-off $\Lambda$. A priori, the possible mass terms in dynamical mass generation are:
\[\frac{1}{2} M^{2}\phi^{2} ,\  \  m\bar{\psi } \psi ,\  i\mu \bar{\psi } \gamma^{5} \psi \]

In order to make analytic progress in understanding dynamical mass generation we will need to make simple ans$\ddot{a}$tze:. To lowest order approximation vertex corrections are neglected. This approximation is called the rainbow approximation.

\begin{eqnarray}
\label{A10}
S^{- 1}_s (k) &=& - i k^2\\
S^{- 1}_f (k) &=& - i \slashed{k}\\
G^{- 1}_s (k) &=& - i (k^2 - M^2)\\
G^{- 1}_f (k) &=& - i \left( \slashed{k} - m-i\mu \gamma_{5} \right)\\
\Gamma^{3}(p,k)&=&g\gamma^{5}\\
\Gamma^{4}(p,q,k)&=&i \lambda \label{A10a}
\end{eqnarray}

In the rainbow approximation we can concentrate  on \eqref{A7} and \eqref{A6}. The presence of a quartic scalar coupling $\lambda$ with both Hermitian and non-Hermitian signs is essential for a consistent \cPT-symmetric formulation of our field theory. New possibilities of solutions become possible in the presence of this extra coupling. 

\subsection{Solution in the Rainbow approximation}\label{sec:rainbow}

Although we allow a chiral mass term $\mu$, we will first discuss $\mu\approx 0$ since these solutions may have lower energy than solutions with $\mu \ne 0$, 
according to the arguments given in  \cite{R10}.\footnote{Our discussion~\cite{R4} on the necessity of a quartic scalar self-coupling, for consistency of renormalisation, is still valid in the presence of a chiral mass term.} We will later consider the case $\mu \ne 0$ and associated solutions. On substituting the equations~\eqref{A10}-\eqref{A10a} in \eqref{A6} and \eqref{A7} we meet integrals of the type 
\begin{equation}
\label{A11}
\int \frac{d^4 p}{(2 \pi)^4} \frac{1}{p^2 + \Delta} = \int^{\Lambda}_0
\frac{p^3 d p}{8 \pi^2} \frac{1}{p^2 + \Delta} =
\frac{1}{16 \pi^2} \left( \Lambda^2 - \Delta \ln \left( 1 +
\frac{\Lambda^2}{\Delta} \right) \right).
\end{equation}
It simplifies the appearance of the equations if we introduce the parameters: 
\begin{align}\label{h} 
h=\frac{g^{2}}{16\pi^{2} }\, ,
\end{align}
where $g$ is the Yukawa coupling (and so $h$ can be positive for $g$ real, or negative for $g$ pure imaginary); let
\begin{align}\label{u}
\tilde{u}=\frac{\lambda}{16\pi^{2}} \,,
\end{align}
where $\lambda$ is the quartic coupling, which can also be positive or negative, and let the dimensionless mass ratios be
\begin{align}\label{r}
r\equiv \frac{M^{2}}{\Lambda^{2} } \ge 0 \,; 
\end{align}
\begin{align}\label{s}
s\equiv\frac{m^{2}}{\Lambda^{2} }\ge 0 \,; 
\end{align}
\begin{align}\label{t}
t\equiv \frac{\mu^{2}}{\Lambda^{2} }\ge 0\,.
\end{align}

From \eqref{A6} we deduce two equations, the first being
\begin{equation}
r-s-t=h\left( r\rm ln \left( 1+\frac{1}{r} \right)  -\left( t+s\right)  \rm ln \left( 1+\frac{1}{t+s} \right)  \right) 
\label{A12}
\end{equation}  
and, the second equation, when $t\ne 0$, is
\be
\left( r\rm ln \left( 1+\frac{1}{r} \right)  -\left( t+s\right)  \rm ln \left( 1+\frac{1}{t+s} \right)  \right)=0 . 
\label{A13}
\ee
to avoid inconsistency.
The SD equation for the scalar propagator \eqref{A7}, in the rainbow approximation, has the form 
\be
r=-4h\left( {\frac{1+s+3t}{1+s+t}} -\left( s+3t \right)  \rm ln \left( 1+{\frac{1}{s+t}} \right)  \right)-\frac{\tilde{u}}{2} \left( 1-s\rm ln \left( 1+\frac{1}{s} \right)  \right)  
\label{A14}
\ee
where the quartic scalar coupling appears.
These three equations will be the key equations for this analysis. 

It has been argued elsewhere that $t \approx 0$ for energetic reasons~\cite{R10,R6a}, and so, instead of \eqref{A12}, we will first consider
\be
r-s=h\left( r\rm ln \left( 1+\frac{1}{r} \right)  -s  \rm ln \left( 1+\frac{1}{s} \right)  \right) 
\label{A15}
\ee
and also, instead of \eq{A14}, we will consider
\be \label{A16}
r=-\left(4h+\frac{\tilde{u}}{2} \right)\left( 1 - \left(s  \rm ln \left( 1+{\frac{1}{s}} \right)  \right)  \right). 
\ee
We recall that: $h<0$ and $\tilde{u}>0$ are both non-Hermitian values;~$h>0$ and $\tilde{u}<0$ are Hermitian values.
\begin{enumerate}
  \item Suppose $s \approx 0$. From \eq{A16} we deduce that $r=-\left(4h+\frac{\tilde{u}}{2}\right)$, and so for $r>0$, 
  \be
   \label{A17}
  h<-\frac{\tilde{u}}{8}\,.
  \ee
 From \eq{A15}, since $r\neq 0$, 
  we obtain 
  \begin{align}\label{M}
  \frac{1}{h} =\rm ln \left( 1+\frac{1}{r} \right) \,,
  \end{align} 
  and 
  \be
  \label{A18}
  h=-\frac{\tilde{u} }{8} -\frac{1}{4\left( \exp \left( \frac{1}{h} \right)  -1\right)  } \,,
  \ee
Equations \eq{A17} and \eq{A18} are compatible.

Moreover, by making the reasonable assumption $\Lambda \gg M^2$, \eqref{M} yields a non-perturbative 
solution for the dynamical pseudoscalar mass, for weak $g$ couplings 
\begin{align}
M^2 \simeq \Lambda^2 \, \exp\Big(-\frac{16\, \pi^2}{g^2}\Big) \, \ll \,  \Lambda^2\, . 
\end{align}

  \item For $r=s$  \eq{A15} is trivially satisfied.
  \eq{A16}, on writing $a\equiv 4h+\frac{\tilde{u} }{2}$, gives 
   \be
    \label{A18B}
   \frac{s}{1-s\rm ln \left( 1+\frac{1}{s} \right)  } =-a
   \ee
   and so $a<0$. This is possible for $g$ non-Hermitian  and $u$ Hermitian with $\left| u\right| $ small.
   
   For $|g|$ also small, that is $0 < -a \ll 1$, we may obtain analytic solutions for $ 0 < r \simeq s \ll 1$, that is dynamical fermion masses, of the form
   \begin{align}\label{m}
   M^2= m^2 = -a\, \Lambda^2 \,   \ll \, \Lambda^2\,.
   \end{align}
  
  \item For $r=0$ and $s \ne 0$, \eq{A14} and \eq{A15} become
  \be
   \label{A19}
  1=h\  \rm ln \left( 1+\frac{1}{s} \right) 
  \ee
  and 
 
  \be
   \label{A20}
  a\  \left( 1-s\rm ln \left( 1+\frac{1}{s} \right)  \right)  =0.
  \ee
\end{enumerate}
For $a\ne 0$ these equations are not compatible since they imply $s=h$ and $s=\frac{1}{\exp \left( \frac{1}{h} \right)  -1} $.There is no $h$ which satisfies both these equations. In the special case that $a=0$ we just have  
  \be
   \label{A21}
   s= \frac{m^2}{\Lambda^2} = \frac{1}{\exp \left( \frac{1}{h} \right)  -1}\, >\, 0\,,
   \ee
    and so given an $h$ we need $\tilde u=-8h$; so non-Hermitan $g$ is associated with a non-Hermitian $\lambda$.
    From \eqref{A21}, and small $|g^2| \ll 1 $ (and, thus, $|\tilde u| \ll 1$), we observe that for Hermitian $g$ and $\tilde u$, 
    we obtain, analytically, non-perturbative fermion masses
 \begin{align}\label{mherm}
 m^2 \simeq  \Lambda^2    \, \exp\Big(-\frac{16\,\pi^2}{|g^2|}\Big) \ll \Lambda^2, \quad   0 < g^2 = - 2\pi^2 \, \tilde u \ll 1\,,
\end{align}
while there is no consistent solution $s>0$ for non-Hermitian $g^2 <0$, $|g^2 | \ll 1$. 

We now come to a discussion for the generation of a chiral mass $\mu$ for fermions. In \cite{R10}, we have given arguments that in the anti-Hermitian Yukawa interaction case, such a generation would not be energetically favourable.

For \emph{nonzero} $t$ we can replace \eq{A12} with
\be
\label{A22}
t=r-s.
\ee
We still have \eq{A13} and \eq{A14}. In \eq{A14}, on using \eq{A22}, we find
\be
\label{A23}
r=-4h\left( \frac{1+3r-2s}{1+r} -\left( 3r-2s\right)  \rm ln \left( 1+\frac{1}{r} \right)  \right)  -\frac{\tilde u}{2} \left( 1-s\rm ln \left( 1+\frac{1}{s} \right)  \right).  
\ee
\begin{itemize}
  \item Seek a solution with $s=0$. From \eq{e11} we deduce that 
  \be
  \label{A24}
  r+12h+\frac{\tilde{u} }{2} =4h\left( \frac{2}{1+r} +3r\rm ln \left( 1+\frac{1}{r} \right)  \right).  
  \ee
  The function $\left( \frac{2}{1+r} +3r\rm ln \left( 1+\frac{1}{r} \right)  \right)$ is positive and greater than 2 for $r$ positive and single-humped. The left-hand side of \eq{A24} is a straight line as a function of $r$. For $h>0$ and  $12h+\frac{\tilde{u} }{2} $ positive and not too large there are two solutions of \eq{A24}. If $12h+\frac{\tilde{u} }{2} $ is negative with $h$ positive there is one solution for $r$. For $h$ negative and $12h+\frac{\tilde{u} }{2}$ positive there are no solutions for $r$. A graphical analysis of \eq{A24} yields in a similar way all possible solutions for different parameter regimes (Hermitian, as well as non-Hermitian). For these solutions $t=r$. 
  This non-zero pseudoscalar mass is consistent with the considerations in \cite{R10}.

  \item A solution with $r=0$ is not allowed since we would then have $t=-s$ which is only compatible with no mass generation. 
   This  is also consistent with the considerations in \cite{R10}.

\end{itemize}

\subsection{Beyond the rainbow approximation: the potential r\^ole of the wave-function renormalisation}\label{sec:beyondrainbow}

 So far, we have ignored the wave function renormalisation, as a first approximation which is not inconsistent with the assumed perturbative nature of the involved couplings. In this section we will consider the effect of including wave function renormalisation, the possible existence of a critical Yukawa coupling for fermion mass generation and alternative solutions to the mass function in the presence of scalar self-interactions, discussed in the model of \cite{bashir}. 

Our starting point is again the SD equations \eqref{A7}, \eqref{A6}, \eqref{A8}, \eqref{A9}. In what follows we 
concentrate on the one-loop SD equations for the propagators of the fermions and psudoscalar fields, respectively, which we give again below for the reader's convenience:
\begin{equation}\label{GF}
  G^{- 1}_f (k) - S^{- 1}_f (k) = - \int_p g \gamma^5 G_f (p) \Gamma^{(3)} (p,
  k) G_s (p - k)
\end{equation}
\begin{equation}\label{GS}
  G^{- 1}_s (k) - S^{- 1}_s (k) = \tmop{tr} \left[ \int_p g \gamma^5 G_f (p)
  \Gamma^{(3)} (p, k) G_f (p - k) \right] - i \frac{\lambda}{2} \int_p G_s (p)
\end{equation}

Incorporation of the wave function renormalisation for the fermion and pseudoscalars (denoted by $F(k^2), S(k^2)$, respectively) means we will use the following form for the dressed propagators
\begin{align}\label{props}
G_f (k) = i \frac{F (k^2)}{\slashed{k} -\mathcal{M} (k^2)}, \quad
G_s (k) = i \frac{S (k^2)}{k^2 -\mathcal{M}_s (k^2)^2}
\end{align}
and we use $\Gamma^{(3)} = g \gamma^5
\Gamma_A$, for the Yukawa vertex. 

With these choices, after some straightforward manipulations, the propagator SD equations \eqref{GF}, \eqref{GS}, can be 
written as
\begin{eqnarray}
\slashed{k} (1 - F (k^2)) -\mathcal{M} (k^2) &=& i g^2 \int_p \gamma^5 \frac{F
(k^2) F (p^2)}{\slashed{p} -\mathcal{M} (p^2)} \gamma^5 \Gamma_A (p, k) \frac{S
((k - p)^2)}{(k - p)^2 -\mathcal{M}_s ((k - p)^2)^2}\\
k^2 (1 - S (k^2)) -\mathcal{M}_s (k^2)^2 &=& - i g^2 \tmop{tr} \left[ \int_p
\gamma^5 \frac{S (k^2) F (p^2)}{\slashed{p} -\mathcal{M} (p^2)} \gamma^5 \Gamma_A
(p, k) \frac{F ((p - k)^2)}{\slashed{p} - \slashed{k} -\mathcal{M} ((p - k)^2)}
\right] \nonumber\\
& & + \frac{\lambda}{2} \int_p i \frac{S (k^2) S (p^2)}{p^2 -\mathcal{M}_s^2
(p^2)}
\end{eqnarray}

We expect that perturbatively $F = 1 +\mathcal{O} (g^2)$, $S = 1 +\mathcal{O}
(\lambda, \, g^2)$ and $\Gamma_A = 1 +\mathcal{O} (g^2)$.\footnote{These conclusions need to be modified, of course, if the analysis is done in the neighborhood of non-trivial infrared fixed points. A full analysis of the fixed points for the Yukawa  theory at two loops or more, incorporating anomalous dimensions,  is then required.} So,  to order
$g^2$ or $\lambda$, ignoring $\mathcal O(g^2\, \lambda)$ and higher order terms, which suffices if we consider perturbatively small couplings,  as we do here, we obtain, after standard manipulations:
\begin{eqnarray}
\slashed{k} (1 - F (k^2)) -\mathcal{M} (k^2) &=& i g^2 \int_p \gamma^5
\frac{1}{\slashed{p} -\mathcal{M} (p^2)} \gamma^5 \frac{1}{(k - p)^2
-\mathcal{M}_s ((k - p)^2)^2} \nonumber \\
&=& i g^2 \int_p  \frac{- \slashed{p}
+\mathcal{M} (p^2)}{(p^2 -\mathcal{M} (p^2)^2) ((k - p)^2 -\mathcal{M}_s ((k -
p)^2)^2)}  \label{eqF} \\
k^2 (1 - S (k^2)) -\mathcal{M}_s (k^2)^2 &=& - i g^2 \tmop{tr} \left[ \int_p
\gamma^5 \frac{1}{\slashed{p} -\mathcal{M} (p^2)} \gamma^5 \frac{1}{\not{p} -
\slashed{k} -\mathcal{M} ((p - k)^2)} \right]  + i \frac{\lambda}{2} \int_p \frac{1}{p^2 -\mathcal{M}_s (p^2)^2} \nonumber \\
&=& - i g^2 \tmop{tr} \left[ \int_p 
\frac{\left( - \slashed{p} +\mathcal{M} (p^2) \right) \left( \slashed{p} - \slashed{k}
+\mathcal{M} ((p - k)^2) \right)}{(p^2 -\mathcal{M}(p^2)^2) ((p - k)^2
-\mathcal{M} ((p - k)^2)^2)} \right]
 + i \frac{\lambda}{2} \int_p \frac{1}{p^2
-\mathcal{M}_s (p^2)^2} \nonumber \\ \label{eqS} 
\end{eqnarray}

The rainbow approximation implies setting the wavefunction renormalisation functions $F$ and $S$ to unity, and assuming constant mass functions, which produced the results in the previous section. 

However, on retaining to this order the $F$ and $S$ leads to a different approach altogether, distinct from the rainbow approximation, 
as we now proceed to demonstrate. Indeed, 
on multiplying \eqref{eqF},
 by $\slashed{k}$, and taking the trace, we obtain, after some  re-arrangements :
\begin{equation}
\label{eqwaveF}
F (k^2) = 1 + i g^2 \int_p  \frac{k \cdot p}{k^2 (p^2 -\mathcal{M}(p^2)^2)
((k - p)^2 -\mathcal{M}_s ((k - p)^2)^2)}\,,
\end{equation}
while, on taking the trace in \eqref{eqF}, yields:
\begin{equation}
\label{eqmassM}
\mathcal{M} (k^2) = - i g^2 \int_p  \frac{\mathcal{M} (p^2)}{(p^2
-\mathcal{M} (p^2)^2) ((k - p)^2 -\mathcal{M}_s ((k - p)^2)^2)}
\end{equation}

The scalar equation \eqref{eqS}, on the other hand, can be manipulated to give:
\begin{eqnarray}
\label{eqwaveS}
S (k^2) &=& 1 - \frac{\mathcal{M}_s (k^2)^2}{k^2} + 4 i g^2 \int_p  \frac{- p^2
+ k \cdot p +\mathcal{M} (p^2) \mathcal{M} ((p - k)^2)}{k^2 (p^2
-\mathcal{M}(p^2)^2) ((p - k)^2 -\mathcal{M}((p - k)^2)^2)} \nonumber\\
& & - i \frac{\lambda}{2}
\int_p \frac{1}{k^2 (p^2 -\mathcal{M}_s (p^2)^2)}
\end{eqnarray}

Upon performing a Wick rotation in the momenta, and doing the angular integrations, 
 the equations \eqref{eqwaveF}, \eqref{eqmassM} and \eqref{eqwaveS} give:
\begin{eqnarray}
F (k^2) &=& 1 - \frac{2 g^2}{(2 \pi)^3 k^2} \int^{\Lambda}_0 \frac{p^3 d p}{p^2
+\mathcal{M} (p^2)^2} \int^{\pi}_0 d \theta \frac{\sin^2 \theta k p \cos
\theta}{p^2 + k^2 - 2 k p \cos \theta +\mathcal{M}_s ((k - p)^2)^2} \label{eqFeuc}\\
\mathcal{M} (k^2) &=& \frac{2 g^2}{(2 \pi)^3} \int^{\Lambda}_0
\frac{\mathcal{M} (p^2) p^3 d p}{p^2 +\mathcal{M} (p^2)^2} \int^{\pi}_0 d
\theta \frac{\sin^2 \theta}{p^2 + k^2 - 2 k p \cos \theta +\mathcal{M}_s ((k -
p)^2)^2} \label{eqMeuc}\\
S (k^2) &=& 1 + \frac{\mathcal{M}_s (k^2)^2}{k^2} + \frac{8 g^2}{(2 \pi)^3 k^2}
\int^{\Lambda}_0 \frac{p^3 (p^2 +\mathcal{M} (p^2) \mathcal{M} ((p - k)^2)) d
p}{p^2 +\mathcal{M} (p^2)^2} \nonumber\\
& & \times \int^{\pi}_0 d \theta \frac{\sin^2 \theta}{p^2 +
k^2 - 2 k p \cos \theta +\mathcal{M} ((p - k)^2)^2} \nonumber\\
& & - \frac{8 g^2}{(2 \pi)^3 k^2} \int^{\Lambda}_0 \frac{p^3 d p}{p^2
+\mathcal{M} (p^2)^2} \int^{\pi}_0 d \theta \frac{\sin^2 \theta k p \cos
\theta}{p^2 + k^2 - 2 k p \cos \theta +\mathcal{M} ((p - k)^2)^2} \nonumber\\
& & + \frac{\pi \lambda}{2 (2 \pi)^3 k^2} \int^{\Lambda}_0 \frac{p^3 d p}{p^2 +\mathcal{M}_s
(p^2)^2} \label{eqSeuc}
\end{eqnarray}

To make progress towards an analytic solution of the above equations, we make the assumption 
\begin{align}\label{smallmass}
\mathcal{M}(p^2)^2  \ll p^2 \quad {\rm  and} \quad  \mathcal{M}_s(p^2)^2  \ll p^2~,
\end{align}
which, we stress, are valid only in the Euclidean space of the Wick rotated momenta. 

Thus, in the following we neglect terms quadratic in the mass
functions. This appears consistent with \eqref{eqMeuc}, because of the presence of the factor $g^2 \ll 1$ on the right-hand side. Therefore, Eqs.  \eqref{eqFeuc}, \eqref{eqMeuc} and \eqref{eqSeuc} read:
\begin{eqnarray}\label{fms}
F (k^2) &=& 1 - \frac{2 g^2}{(2 \pi)^3 k^2} \int^{\Lambda}_0 p d p \int^{\pi}_0
d \theta \frac{\sin^2 \theta k p \cos \theta}{p^2 + k^2 - 2 k p \cos \theta} \,,\nonumber\\
\mathcal{M} (k^2) &=& \frac{2 g^2}{(2 \pi)^3} \int^{\Lambda}_0 \mathcal{M}
(p^2) p d p \int^{\pi}_0 d \theta \frac{\sin^2 \theta}{p^2 + k^2 - 2 k p \cos
\theta}\,, \nonumber\\
S (k^2) &=& 1 + \frac{8 g^2}{(2 \pi)^3 k^2} \int^{\Lambda}_0 p^3 d p
\int^{\pi}_0 d \theta \frac{\sin^2 \theta}{p^2 + k^2 - 2 k p \cos \theta} \nonumber\\
& & - \frac{8 g^2}{(2 \pi)^3 k^2} \int^{\Lambda}_0 p d p \int^{\pi}_0 d \theta
\frac{\sin^2 \theta k p \cos \theta}{p^2 + k^2 - 2 k p \cos \theta} \nonumber\\
& & + \frac{\pi \lambda}{2 (2 \pi)^3 k^2} \int^{\Lambda}_0 p\, d p\,.
\end{eqnarray}

Using
\begin{align}
&\int^{\pi}_0 d \theta \frac{\sin^2 \theta}{(p^2 + k^2 - 2 k p \cos \theta)}
= \frac{\pi}{2} \left( \frac{1}{p^2} \theta (p^2 - k^2) + \frac{1}{k^2} \theta
(k^2 - p^2) \right), \quad {\rm and} \nonumber \\
&\int^{\pi}_0 d \theta \frac{\sin^2 \theta k p \cos \theta}{(p^2 + k^2 -
2 k p \cos \theta)} = \frac{\pi}{4} \left( \frac{k^2}{p^2} \theta (p^2 - k^2)
+ \frac{p^2}{k^2} \theta (k^2 - p^2) \right)\,,\nonumber 
\end{align}
and changing variables appropriately, we obtain from \eqref{fms}:
\begin{eqnarray}
F (k^2) &=& 1 - \frac{\pi g^2}{4 (2 \pi)^3 k^2} \int^{\Lambda^2}_0 d p^2 \left(
\frac{k^2}{p^2} \theta (p^2 - k^2) + \frac{p^2}{k^2} \theta (k^2 - p^2) \right) \label{eqftemp}\\
\mathcal{M} (k^2) &=& \frac{\pi g^2}{2 (2 \pi)^3} \int^{\Lambda^2}_0
\mathcal{M} (p^2) d p^2 \left( \frac{1}{p^2} \theta (p^2 - k^2) +
\frac{1}{k^2} \theta (k^2 - p^2) \right) \label{eqmtemp}\\
S (k^2) &=& 1 + \frac{g^2}{(2 \pi)^2 k^2} \int^{\Lambda^2}_0 p^2 d p^2 \left(
\frac{1}{p^2} \theta (p^2 - k^2) + \frac{1}{k^2} \theta (k^2 - p^2) \right) \nonumber\\
& & - \frac{g^2}{2 (2 \pi)^2 k^2} \int^{\Lambda^2}_0 d p^2 \left( \frac{k^2}{p^2}
\theta (p^2 - k^2) + \frac{p^2}{k^2} \theta (k^2 - p^2) \right) \nonumber\\
& & + \frac{\pi \lambda}{4 (2 \pi)^3 k^2} \int^{\Lambda^2}_0 d p^2 \label{eqstemp}
\end{eqnarray}

Doing the (Euclidean) momentum integrals in \eqref{eqftemp} and \eqref{eqstemp}, we obtain:
\begin{eqnarray}
F (k^2) &=& 1 - \frac{g^2}{32 \pi^2} \ln \left( \frac{\Lambda^2}{k^2} \right) -
\frac{g^2}{64 \pi^2}\\
S (k^2) &=& 1 - \frac{g^2}{8 \pi^2} \ln \left( \frac{\Lambda^2}{k^2} \right) +
\frac{1}{4 \pi^2} \left( g^2 + \frac{\lambda}{8} \right) \left(
\frac{\Lambda^2}{k^2} \right) - \frac{3 g^2}{16 \pi^2}
\end{eqnarray}

The equation for $F$ is the same as in \cite{bashir}, upon keeping the leading
logarithm. The difference between our result and the result of \cite{bashir} lies in the fact that we have introduced a wavefunction renormalization for the scalar $S (k^2)$, as well.

Now we focus on the mass function equation \eqref{eqmtemp}. If we take the  ansatz 
for the mass function in the massless phase of the theory, 
\begin{align}\label{mps}
\mathcal{M} (p^2) = m_0 \, \Big(\frac{p}{m_0}\Big)^{- 2 s}\,, \quad s > 0\,,
\end{align}
(as in~\cite{bashir}) with the quantity $s$ to be determined, where $m_0$ is a mass scale to be determined below. Naively, 
since the cutoff $\Lambda$ plays the r\^ole of the only mass scale in the (bare) system, one would be tempted to identify 
$m_0 = \Lambda $. However, given that the validity of the effective theory requires $p < \Lambda$ in Euclidean momentum space, this identification would be inconsistent with \eqref{mps}. The mass scale $m_0$ should then be identified with some other infrared scale of the theory, such that $p \gg m_0$. Such a scale could be provided by the dynamically generated (fermion) mass. In such a case, \eqref{mps} would be consistent with the assumption
\eqref{smallmass}, allowing for an analytic treatment of the SD equations, provided 
\begin{align}\label{suff}
\Big(\frac{p^2}{m^2_0}\Big)^{\frac{1}{2}+ s} \, \gg \, 1\,  \quad s > 0\,, \quad \Rightarrow \quad  p \gg m_0 ~,   
\end{align}
where the latter condition is a sufficient condition, consistent with $m_0$ being an infrared (IR) scale in the problem, perhaps arising dynamically (or through, e.g. curved space time effects, such as the cosmological vacuum energy). 

Let us examine the consistency of this approach. 
Upon using \eqref{mps}, Eq.~\eqref{eqmtemp} becomes:
\begin{equation}\label{finaleq}
k^{- 2 s} = \frac{g^2}{16 \pi^2} \left( \int^{\Lambda^2}_{k^2} (p^2)^{- (s +
1)} d p^2 + \frac{1}{k^2} \int^{k^2}_0 (p^2)^{- s} d p^2 \right)\, 
\end{equation}
which, upon integration, yields:\footnote{Technically, the lowest bound on the $p$ integration should be $m_0 \ll \Lambda$. Such contributions, however, are negligible compared to the rest of the terms in \eqref{finaleq}, and hence 
one can safely let $m_0 \to 0$ for the purposes of manipulating this equation.}
\begin{equation}\label{crit}
1 = \frac{g^2}{16 \pi^2} \left( \frac{1}{s (1 - s)} - \frac{1}{s} \left(
\frac{k}{\Lambda} \right)^{2 s} \right)\, 
\end{equation}
where convergence of the integral at $0$ requires $s < 1$.
Technically this equation is inconsistent, as the right-hand side depends on the momenta, while the 
left hand side does not.  One, however, may assume the validity of this equation in the regime 
({\it cf.} \eqref{smallmass}) that $\mathcal{M} (k^2)^2 \ll k^2 \ll \Lambda^2$ 
in which case one can neglect $\left( \frac{k}{\Lambda} \right)^{2 s}$ in \eqref{crit}. Thus, upon this {\it approximation}, 
we obtain
\begin{equation}\label{ssol}
1 = \frac{g^2}{16 \pi^2 s (1 - s)} \qquad \Rightarrow \qquad s = \frac{1}{2} \pm
\frac{1}{2} \sqrt{1 - \frac{g^2}{4 \pi^2}}\,,
\end{equation}
with $ 1 > s > 0$ consistent with the initial assumption \eqref{mps}, provided the quantity in the square root is non negative, that is $g^2 \le 4\pi^2$. This is guaranteed in our case, due to the perturbative assumption 
of small couplings $g^2 \ll 1$, in which case the two solutions for $0 < s < 1$ read :
\begin{align}\label{scalexp}
s_+ \simeq 1 - \frac{g^2}{8\pi^2},  \quad s_- \simeq \frac{g^2}{8\pi^2}, \quad g^2 \ll 1\,.
\end{align}
In this case, there is no dynamical mass generation, given the form of the mass function \eqref{mps}, which diverges as $p \to 0$.

We note, for completeness, that the ansatz \eqref{mps}   and the existence of the scaling 
exponents \eqref{scalexp} follow rigorously,
on using a different method of solution of \eqref{eqmtemp} (which is derived from \eqref{eqMeuc} and \eqref{smallmass}). In particular we regard \eqref{smallmass} as an asymptotic condition for large $p$. In order to analyse \eqref{eqmtemp}, we derive from it a second order differential equation, which is then solved. 

Let us introduce the parameters $y=\frac{p^{2}}{m^{2}_{0} } ,\  z=\frac{k^{2}}{m^{2}_{0} } \  ,\  \  \tilde{M} \left( z\right)  =\frac{M\left( z\right)  }{m_{0} }, \kappa =\frac{\Lambda^{2} }{m^{2}_{0}}  $. In terms of these variables \eqref{eqmtemp} becomes
\begin{equation}
\label{S1}
\tilde{M} \left( z\right)  =\alpha \  \int^{\kappa}_{1} dy\  \tilde{M} \left( y\right)  \left( \frac{1}{y} \Theta \left( y-z\right)  +\frac{1}{z} \Theta \left( z-y\right)  \right)  
\end{equation}
where $\alpha \  \equiv \frac{\pi g^{2}}{2\  \left( 2\pi \right)^{3}  } $. \eqref{S1} implies the second order differential equation
\begin{equation}
\label{S2}
\frac{d}{dz} \left( z^{2}\frac{d}{dz} \tilde{M} \left( z\right)  \right)  =-\alpha \  \  \tilde{M} \left( z\right) . 
\end{equation}
and the boundary condition

\begin{equation}
\label{S3}
  \tilde{M} \left( \kappa\right)    = \frac{\alpha}{\kappa} \int^{\kappa}_{1} dy\  \tilde{M} \left( y\right).   
\end{equation}

The solution of \eqref{S2}, on applying \eqref{S3}, is
\begin{equation}
\label{S4}
\tilde{M} \left( z\right)=c_1 z^{\frac{1}{2}
   \left(-\sqrt{1-4 \alpha
   }-1\right)}+c_2
   z^{\frac{1}{2}
   \left(\sqrt{1-4 \alpha
   }-1\right)}
\end{equation}
with $c_2=-\frac{c_1 \kappa
   ^{\frac{1}{2}
   \left(-\sqrt{1-4 \alpha
   }-1\right)}
   \left(\left(\sqrt{1-4
   \alpha }+1\right) \kappa
   ^{\frac{1}{2}
   \left(\sqrt{1-4 \alpha
   }+1\right)}-\sqrt{1-4
   \alpha } \kappa +\kappa
   \right)}{\left(\sqrt{1-4
   \alpha }+1\right) \kappa
   ^{\frac{1}{2}
   \left(\sqrt{1-4 \alpha
   }+1\right)}-\sqrt{1-4
   \alpha }+1}$. We require $1>4 \alpha $ which is always possible for non-Hermitian $\alpha$ and upto a critical value for Hermitian $\alpha$. For $\alpha >1/4$ the mass function takes complex values, indicating the possibility of a  phase transition. The study of such a phase, however, requires going beyond the one-loop SD approximation and a numerical treatment. In the model of \cite{bashir}, which has a specific scalar self-interaction coupling proportional to $g^2$ and ignores the dynamical generation in the (pseudo)scalar sector, 
such a treatment, in the phase where $g > g_c$, leads to a numerical fit for the fermion mass  of the form 
  \begin{align}\label{mnumer}
M = \Lambda \, \Big( \exp(- \frac{\mathcal A}{\sqrt{\frac{g^2}{g_c^2} -1 }} + \mathcal B\Big) \,,
\end{align} 
where $\mathcal A, \, \mathcal B > 0$ of order $\mathcal O(1)$.  
In our model, which also involves pseudoscalar fields with self-interaction couplings independent of $g$, it will be necessary to perform a more complicated analysis to study dynamical mass generation for both fermion and pseudoscalar fields, taking into account any renormalisation group infrared fixed points. 

 We now discuss a way to recover the 
constant dynamical mass generation of the rainbow approximation given earlier.
 By considering 
 a constant fermion mass function,  
$\mathcal{M} (p^2) = m$, in \eqref{eqmtemp} where 
 $m$ is considered to be a very small quantity (compared to the energy scales in the problem), we can self-consistently neglect terms of order $m^2$. In this case \eqref{eqmtemp}, would {\it naively} become:
\begin{equation}
1 = \frac{g^2}{16 \pi^2} \left( \int^{\Lambda^2}_{k^2} \frac{1}{p^2} d p^2 +
\frac{1}{k^2} \int^{k^2}_0 d p^2 \right)\, 
\end{equation}
which, upon integration, gives:
\begin{equation}
g^2 = \frac{16 \pi^2}{1 - \ln \left( \frac{k^2}{\Lambda^2} \right)}.
\end{equation}
This is not quite consistent, as it  implies a momentum dependent coupling $g^2$. 
 In view of \eqref{smallmass},  the lower bound of the $p$ integration in \eqref{eqmtemp} should be the small, dynamically generated, fermion mass function $\mathcal M(k^2)$ itself, which would serve as an IR cutoff. In principle this would turn \eqref{smallmass} into an integral equation, which is difficult to solve analytically. Nonetheless, in the IR limit, where the external momenta $k^2 $ tend to an IR cutoff, provided by a constant fermion mass function, i.e. $k^2 \to m^2 \ll \Lambda^2$, we consider
\begin{align}\label{IR}
\mathcal M^2 (k^2=m^2) = m^2 = {\rm constant} > 0\,,
\end{align}
From the modified \eqref{eqmtemp},  where the lower bound of the $p$ integration is identified with $\mathcal M(k^2) \ll \Lambda$, evaluated at the IR limit $k^2=m^2$:
\begin{equation}\label{improvedM}
\mathcal{M} (k^2)\,\Big |_{k^2=m^2}  \simeq \frac{g^2}{16\pi^2} \Big[\int^{\Lambda^2}_{k^2} 
\mathcal{M} (p^2) d p^2 \, \frac{1}{p^2} + \frac{1}{k^2}\, \int^{k^2}_{\mathcal M(k^2)} \,  
\mathcal{M} (p^2) \, d p^2 \, \Big]\,\Big |_{k^2=m^2} \,.
\end{equation}
Assuming that in this regime, the dominant contribution to the integral on the right-hand-side is obtained from $\mathcal M(p^2=m^2)=m^2 = {\rm constant}$ ({\it cf.} \eqref{IR}),
we easily obtain, upon performing the momentum integrals:
\begin{align}\label{solfinal}
m^2 \simeq \Lambda^2 \, \exp\Big(-\frac{16\, \pi^2}{g^2}\Big)\, \ll \, \Lambda^2\, , \quad g^2 > 0\, ,
\end{align}
which is the dynamically-generated fermion mass in the rainbow approximation for the Hermitian  interactions case \eqref{mherm}. Again, we find that there is no consistent solution for dynamical fermion mass $m < \Lambda$ in the non-Hermitian Yukawa case $g^2 < 0$. 
This demonstrates the validity of the main results for the dynamical fermion mass obtained in the rainbow approximation in this case, 
even if one considers the effects of the wave-function renormalization. This is one potential solution, in the phase diagram of the theory for perturbative $|g|^2 \ll 1$. In the strongly coupled regime of the phase diagram of the system, where the Yukawa coupling is above a critical value, the dynamical fermion mass 
\eqref{mnumer} provides an alternative non-perturbative solution.

\section{Conclusions}\label{sec:concl}

The Lagrangian studied here is of importance for understanding the role of Kalb-Ramond axion in a host of situations such as leptogenesis, dark matter and the strong CP problem~\cite{R15, R16}. It also provides a rationale for a simple \cPT symmetric renormalisable model which can be understood using field theoretic methods. We have shown~\cite{R4} how non-Hermiticity in a renormalisable field theory with a fermion and KR axion is expressed in a path integral formulation at the level of the bare Lagrangian and at the level of the renormalised Lagrangian. Because we allow for non-Hermitian interactions which are \cPT-symmetric, some issues in applying conventional field theory methods arise. These issues are discussed in Appendices.  In quantum mechanics \cPT symmetry is enough to guarantee a unitary theory but going from a finite to an infinite number of degrees of freedom renders the path integral measure nontrivial. The requirement of renormalisation is a prime example of this nontriviality.~Recently~\cite{R4} we have shown that obtaining a perturbative formulation with associated Feynman rules is feasible. However the study of  \cPT symmetry also requires a nonperturbative approach. One way that this can be seen is in the context of a semiclassical analysis of path integrals for \cPT field theories where contributions of trivial and non-trivial saddle points conspire together to make quantities such as the ground state energy finite\footnote{See a recent work \cite{R14a}.}.The SD equations represent one way of going beyond low order perturbation theory based on an expansion around the trivial fixed point. We have considered the SD equations in dimensions $D=0,1$ and  $4$ for theories with no bare mass. In the case of $D=0,1$ we have noted a different mechanism for mass generation that follow from SD analysis for theories which are nonHermitian but \cPT symmetric in the scalar sector. The coupling constant dependence of the mass generation is distinct from that found using SD analysis.

 In $D=4$, we have considered dynamical mass generation using conventional SD equations with a momentum cut-off $\Lambda$ for our  Yukawa theory. The couplings can be Hermitian or non-Hermitian (but \cPT symmetric). The SD equations are considered in two approximations: one the rainbow approximation and the second incorporates wave function renormalisation and goes beyond the rainbow approximation. The rainbow approximation, because of its simplicity,  allows a detailed investigation of dynamical mass generation. 
 
 In the rainbow approximation we found:
 \begin{enumerate}
  \item Only a  non-zero scalar mass is generated for Hermitian couplings.
  \item For the case of equal scalar and (standard, i.e. nonchiral) fermion masses need: non-Hermitian Yukawa coupling but Hermitian quartic scalar coupling;  if the quartic coupling is sufficiently small it can also be allowed to be nonHermitian. 
  \item Only a non-zero (standard) fermion mass can occur if the Yukawa and quartic couplings are both Hermitian.
  \item Equal nonzero pseudoscalar  and chiral fermion masses can arise if Yukawa coupling is Hermitian and quartic scalar coupling is nonHermitian; for sufficiently small quartic coupling (and Hermitian Yukawa coupling) this case is also possible.
\end{enumerate}
In dynamical symmetry breaking, for generation of fermion masses, there can be critical values of couplings below which dynamical symmetry breaking does not arise. The rainbow approximation is too simple to account for this. Taking into account wave function renormalisation (following \cite{bashir}) we obtain evidence for a critical Yukawa coupling for dynamical fermion mass generation. Our analysis differs in two important ways to that in \cite{bashir}: we consider the scalar and fermion wave functions renormalisations and the linearised contributions from the mass functions. We show that within these approximations it is possible to solve the equations without resorting to ansatzes. A critical value of the coupling is a result  of the approximation. However, given that the analysis is based on effectively summing up perturbation theory around the trivial saddle point the conclusion of the existence of a critical coupling should only be regarded as suggestive. A renormalisation group analysis of one particle irreducible two point functions an epsilon expansion may evade this criticism of reliability of (summed-up) perturbation theory , by having a small parameter, epsilon. other than the coupling which can control the size of terms which are ignored.

\section*{Acknowledgments}

We would like to thank Carl Bender and Wen-Yuan Ai for valuable discussions.
The work of N.E.M. and S.S. is supported in part by the UK Science and Technology Facilities research
Council (STFC) and UK Engineering and Physical Sciences Research
Council (EPSRC) under the research grants ST/T000759/1 and  EP/V002821/1, respectively. NEM acknowledges participation in the COST Association Action CA18108 {\it Quantum Gravity Phenomenology in the Multimessenger Approach (QG-MM)}. 
\vspace{.5cm}


\appendix{}

\section{Aspects of Hermiticity\label{sec:appA}}

\subsection{The scalar self-interaction}

 A conventional way of obtaining a potential such as $V_{1}(\phi)=-\frac{\lambda}{4!}\phi^{4}$ is to consider an analytic continuation of the coupling constant  
  $\lambda \to \lambda\exp \left( i\alpha \right) $ which leads to
\begin{equation}
\label{E10}
V_{2}\left( \phi \right)  =\frac{\lambda}{4!}\exp \left( i\alpha \right)  \phi^{4} .
\end{equation}
On starting at $\alpha=0$ and letting $\alpha \to \pi$ (or alternatively $\alpha \to -\pi$) we obtain $V_{1}(\phi)=-\frac{\lambda}{4!}\phi^{4}$. 

A \cPT-symmetric deformation way of obtaining  the same unstable potential is to consider $V_{3}(\phi)=\frac{\lambda}{4!} \phi^{2} \left( i\phi \right)^{\delta }$ obtained from letting  $\delta \to 2$. However (in $D=1$ ) the spectrum of the Hamiltonian with $V_{2}(\phi)$ and and with $V_{3}(\phi)$ differ significantly. This difference is due to the different boundary conditions (encoded in Stokes sectors~\cite{R11}) when calculating the partition function. $V_{2}\left( \phi \right)$ has a spectrum with non-zero imaginary parts.~$V_{3}\left( \phi \right)$ has a spectrum with purely real energy eigenvalues for $\delta \ge 0$.

\begin{figure}[!h]
\centering\includegraphics[width=4.5in]{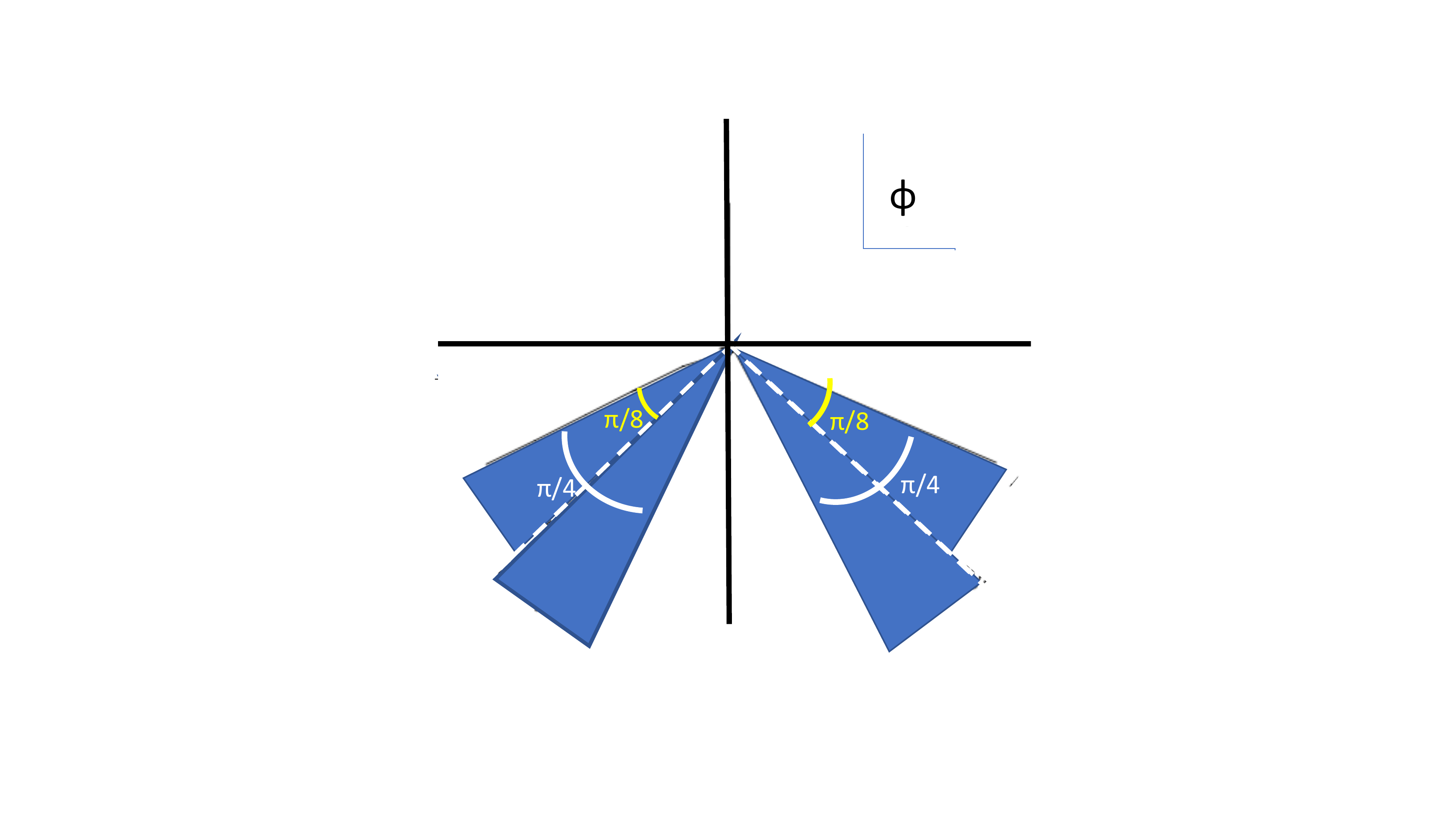}
\caption{ \cPT symmetric Stokes wedges for quartic potential}
\label{fig:diagQuartic}
\end{figure}

We shall start off in the simplest context: bosonic path integrals with discrete $\mathcal P$ and $\mathcal T$ symmetries.
The   (Euclidean) 
 action that will be considered is of the following type  \be
S\left( \varphi \right)  =\int d^{D}x\left( \frac{1}{2} \  \left( \partial_{\mu } \varphi \right)^{2} + \frac{1}{2} m^2\varphi^2  + V^{\rm int} \left( \varphi \right)  \right) . 
\ee

 where $m$ is the mass. The canonical form of $V^{\rm inf} \left( \varphi \right) $ used in the study of \cPT~symmetry is 
 
\be
\label{E11b}
V^{\rm int}\left( \varphi \right)  =  \frac{u}{4!} \varphi^{2} \left( i\varphi \right)^{\delta }  
\ee
with $u$ and $\delta$ real. The action of  \cPT~on $V\left( \varphi \right)$ is determined through:
 \begin{align}\label{cpttrn}
  \mathcal{P}&:\quad \varphi \longrightarrow -\varphi \nonumber \\
  \mathcal{T}&: \quad  \varphi \longrightarrow \varphi \nonumber \\
  \mathcal{T}&:\quad i \longrightarrow -i \,.
\end{align}
The potential  $V\left( \varphi \right)$ is \cPT-symmetric for all values of $\delta$.  For $\delta=2$  we have the negative quartic potential which is conventionally an unstable potential and energies of states have an imaginary part. The above \cPT~symmetric formulation, involving a complex deformation of the potential, leads to a theory in $D=0$ and $D=1$ with a real partition function and real energies respectively. There are strong grounds to expect similar properties to hold for $D>1$. One purpose of this section is to outline the analysis of the integral in $D=0$ in such a way that the generalisation to $D>0$ is clear (but may have complications such as  renormalisation). The path in $\varphi$ space, because of the deformation parametrised by $\delta$,  is required to explore the complex $\varphi$-plane. The presence of \cPT~symmetry results in a left-right symmetry of the Stokes wedges for the deformed path (see Fig. [\ref{fig:diagQuartic}]), i.e. a reflection symmetry in the imaginary $\varphi $-axis. 
This left-right symmetry is responsible for real energy eigenvalues. If, for example,  we have $\mathcal{T}:\  \varphi \longrightarrow -\varphi$ then we do not have \cPT symmetry for general $\delta$, the boundary conditions are different and the left-right symmetry of the deformed paths no longer holds. If the Lagrangian (e.g. for $\delta=2$) formally shows \cPT symmetry for $\mathcal{T}:\  \varphi \longrightarrow -\varphi$ the physical consequences of the different assignments of $\mathcal P$ and $\mathcal T$ are entirely different; one case may give an acceptable physical theory with left-right symmetry and real eigenvalues, while the other case with up-down symmetry would not have real eigenvalues which are bounded below. We will consider below the Euclidean version of the path integral to improve  the convergence of the path integral.
The partition function for $D=0$ has the form
\be
\label{e3}
Z=\int_{C} d\varphi\,_{} \exp\left(-\left(\frac{1}{2}m^2\varphi^2-\frac{1}{4!}u \varphi^4\right)\right).
\ee

$Z$ represents a zero-dimensional field theory~\cite{R2} and the path integral measure is the measure for contour integration. The study of this toy model (which can formally be investigated as a field theory with Feynman rules)  will help in understanding the role of Stokes wedges~\cite{R11} in path integrals.\footnote{
 For a rigorous perspective on the
use of Stokes wedges in complexified path-integrals in quantum mechanics, and in some (related) three-dimensional Chern-Simons gauge theories formulated over complex Lie algebras, see \cite{R11a,R11b}. In our context, the purely imaginary Chern-Simons-axion couplings in section \ref{sec:nhystring} arise for a different reason. Nonetheless, the methods in \cite{R11a,R11b} might be relevant for treating such couplings. We hope to be able to study such issues in the future. } 
  For $u>0$ the integral with the contour $-\infty <\varphi <\infty $ does not exist. For $u<0$ the integral with the contour exists in the Stokes wedges $-\frac{\pi }{8} <\arg \varphi <\frac{\pi }{8} $ and $\frac{7\pi }{8} <\arg \varphi <\frac{9\pi }{8} $. Hence the \emph{conventional} Hermitian theory can use the contour $-\infty <\varphi <\infty $ which goes through the centre of both Stokes wedges. It is straightforward to see that there are $4$ possible Stokes wedges each with an opening of $\pi /4$. 
In a \cPT-symmetric context the partition function can exist for a  contour ${C}$ in the complex $\varphi$-plane, chosen to lie in the  Stokes wedges: $-\frac{3\pi }{8} <\arg \varphi <-\frac{\pi }{8}$ and $-\frac{7\pi }{8} <\arg \varphi <-\frac{5\pi }{8} $. These Stokes wedges are left right symmetric and so the \cPT symmetric theory has real eigenvalues which are bounded below. 

These arguments given explicitly for $D=0$ can be generalised to functional paths or Lefschetz thimbles for $D > 0$.

 \subsection{Fermionic path integrals and their role in \cPT symmetry \label{sec:fermionic}}

\begin{figure}[!h]
\centering\includegraphics[width=4.5in]{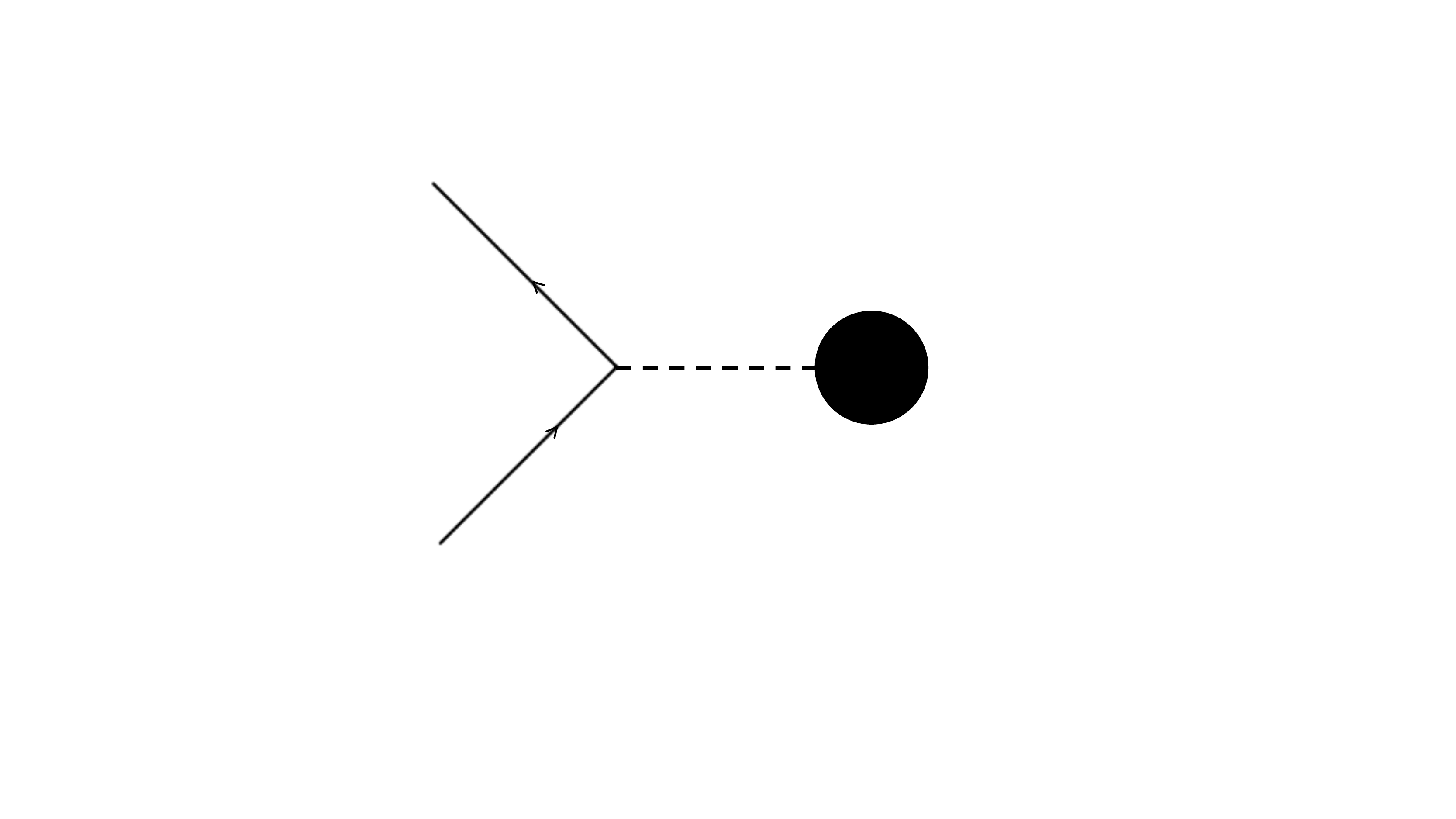}
\vspace{-2.0cm}
\caption{ The master vertex for functional determinant. Continuous lines with arrows denote fermions. The dashed line ending in the dark blob denotes 
an external scalar field source.}
\label{fig:diagVertex}
\end{figure}

\begin{figure}[!h]
\centering\includegraphics[width=4.5in]{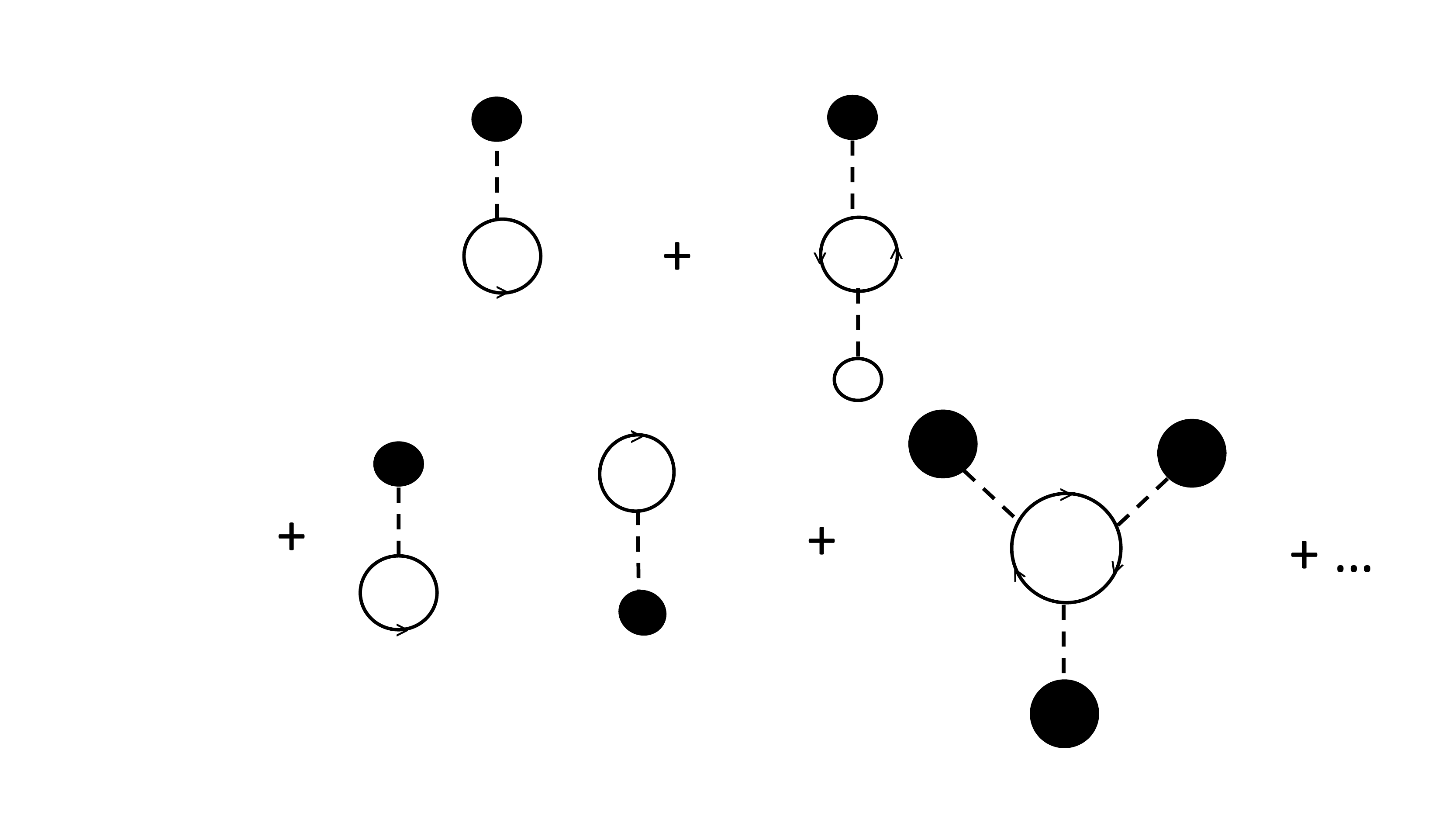}
\caption{Lowest functional vertices for the determinant, including disconnected graphs. The symbols are as in fig.~\ref{fig:diagVertex}.}
\label{fig:diagDeterminant}
\end{figure}

An essential feature of our model is the presence of fermions~\cite{R12}. Since our analysis is based on path integrals we need to check whether the findings on bosonic path integrals are modified by the presence of fermions. The fermionic part of the path integral is in terms of Grassmann numbers which are anticommuting numbers and so Gaussians of Grassmann numbers truncate; at this level there should not be any additional convergence issues in the fermionic theory. To investigate further, since fermions appear quadratically in $L_F$, they can be formally integrated out in the partition function $Z_{eff} $ associated with Eq.\eqref{E1}:
\be
Z_{eff}=\int D\phi \exp \left[ -S_{B}\left( \varphi \right)  \right]  \det \left( \gamma^{\mu } \partial_{\mu } +im+ig\gamma_{5} \varphi \right)  
\ee
where 
\be
\det_{} \left( \gamma^{\mu } \partial_{\mu } +im+ig\gamma_{5} \varphi \right)  =\int D\psi^{\dag } D\psi \exp \left( -\psi^{\dag } \left[  \gamma^{\mu } \partial_{\mu } + im +ig \gamma_5  \varphi \right]  \psi \right).
\ee  
These fermionic determinants have been widely studied using Feynman-diagram representations (see Figs.~\ref{fig:diagVertex} and \ref{fig:diagDeterminant}), and are complicated. 

The formal expressions for these determinants are generally nonlocal; for a heavy fermion mass, these determinants can be approximated~\cite{R13} using semi-classical methods, which leads to a general effective action dependent on the couplings of the fermions prior to the integration~\cite{R14}.~There is an indication that corrections to the bosonic part of the Lagrangian is of of the form $-u^{2}\varphi^{4} $ and $g^{4} \varphi^{4}$. Consequently quantum fluctuations may contribute to non-Hermitian behaviour. However the issues of convergence of the resultant  scalar functional integral can be addressed within the framework of paths in Stokes wedges generalised to Lefschetz thimbles.

\section{Derivation of the Schwinger Dyson Equations \label{sec:appC}}

 In this Appendix, we sketch the details leading to the derivation of the SD equations \eqref{A7}, \eqref{A6}, \eqref{A8}, \eqref{A9}, which we used in the main text to study dynamical mass generation for pseudoscalar and fermion fields in the massless theory \eqref{e53}.
 
Our starting point is the massless Minkowski-space path integral, with partition function $Z$ 
in the presence of appropriate sources, and 
 \begin{eqnarray}
Z [J, \eta, \bar{\eta}] &=& \int \mathcal{D} [\phi \psi \bar{\psi}] \exp
  \left\{ i \, \int d^4 x \left[ \frac{1}{2} \partial_{\mu} \phi \partial^{\mu}
  \phi + \bar{\psi} i \slashed{\partial} \psi - i g \phi \bar{\psi} \gamma^5 \psi
  + \frac{\lambda}{4!} \phi^4 \right] \right.\nonumber\\
& & \left.+  i\, \int d^4 x [J \phi + \bar{\psi} \eta +
  \bar{\eta} \psi] \right\}\label{e53mink}
  \end{eqnarray}

We use the following relation
\begin{equation}
Z=e^{-i W}
\end{equation}
to obtain the propagators:
\begin{eqnarray}
 \frac{\delta^2 W}{\delta J (z) \delta J (x)} &=& - i G_s (x - z) \label{propesc}\\
\frac{\delta^2 W}{\delta \bar{\eta} (z) \delta \eta (x)} &=& i G_f (x -z) \label{propfer}
\end{eqnarray}

We can relate the effective action $\Gamma[\phi,\psi, \bar \psi]$ with this functional $W$ as standard:
\begin{equation}
  W [J, \eta, \bar{\eta}] = - \Gamma [\phi, \psi, \bar{\psi}] - \int d^4 x [J
  \phi + \bar{\psi} \eta + \bar{\eta} \psi]
\end{equation}
Using this relation we get

\begin{equation}
\frac{\delta W}{\delta J} = - \phi,\qquad \frac{\delta W}{\delta \bar{\eta}} = -
\psi \qquad, \frac{\delta W}{\delta \eta} = \bar{\psi}
\end{equation}

and

\begin{equation}
\frac{\delta \Gamma}{\delta \phi} = - J,\qquad \frac{\delta \Gamma}{\delta
\bar{\psi}} = - \eta,\qquad \frac{\delta \Gamma}{\delta \psi} = \bar{\eta}
\end{equation}

From these relations:

\begin{equation}
\frac{\delta^2 W}{\delta \phi (y) \delta J (x)} = - \delta^{(4)} (x - y)
\Rightarrow \int d^4 z \frac{\delta J (z)}{\delta \phi (y)} \frac{\delta^2
W}{\delta J (z) \delta J (x)} = - \delta^{(4)} (x - y)
\end{equation}

\begin{equation}
  \int d^4 z \frac{\delta \Gamma}{\delta \phi (y) \delta \phi (z)}
  \frac{\delta^2 W}{\delta J (z) \delta J (x)} = \delta^{(4)} (x - y)
\end{equation}

Using \eqref{propesc}, we can identify $\frac{\delta^2 \Gamma}{\delta \phi (y) \delta \phi (z)} = i G^{- 1}_s (y -z)$.

In addition,

\begin{equation}
\frac{\delta^2 W}{\delta \bar{\psi} (y) \delta \eta (x)} = \delta^{(4)} (x -
y) \Rightarrow \int d^4 z \frac{\delta \bar{\eta} (z)}{\delta \bar{\psi} (y)}
\frac{\delta^2 W}{\delta \bar{\eta} (z) \delta \eta (x)} = \delta^{(4)} (x -
y)
\end{equation}
\begin{equation}
  \int d^4 z \frac{\delta^2 \Gamma}{\delta \bar{\psi} (y) \delta \psi (z)}
  \frac{\delta^2 W}{\delta \bar{\eta} (z) \delta \eta (x)} = \delta^{(4)} (x -
  y)\,.
\end{equation}

Using \eqref{propfer} we get that $\frac{\delta^2 \Gamma}{\delta \bar{\psi} (y) \delta \psi (z)} = -i G^{- 1}_f (z - y)$.

\subsection{Yukawa fermion-pseudocalar Vertex $\Gamma^{(3)}$}
To get the dressed vertex joining the scalar with fermion and antifermion, we start deriving the fermionic propagator $\frac{\delta^2 W}{\delta \bar{\psi} (y)
\delta \eta (z)} $ respect to $J (x)$:

\begin{equation}
\frac{\delta^3 W}{\delta J (x) \delta \bar{\eta} (y) \delta \eta (z)} =
\frac{i \delta}{\delta J (x)} G_f (z - y) =
\frac{\delta}{\delta J (x)} \left( \frac{\delta^2 \Gamma}{\delta \bar{\psi}
(z) \delta \psi (y)} \right)^{- 1}
\end{equation}

We use that $\frac{d}{dx}M^{-1}(x)=-M^{-1}(\frac{d}{dx}M(x))M^{-1}(x)$ and

\begin{equation}
\frac{\delta^3 W}{\delta J (x) \delta \bar{\eta} (y) \delta \eta (z)} = -
\int d^4 v d^4 w \left( \frac{\delta^2 \Gamma}{\delta \bar{\psi} (z) \delta
\psi (v)} \right)^{- 1} \left( \frac{\delta^3 \Gamma}{\delta J (x) \delta
\bar{\psi} (v) \delta \psi (w)} \right) \left( \frac{\delta^2 \Gamma}{\delta
\bar{\psi} (w) \delta \psi (y)} \right)^{- 1}
\end{equation}

Applying the functional-differentiation chain rule, we introduce the derivative of the source in the third functional derivative:

\begin{align}
\frac{\delta^3 W}{\delta J (x) \delta \bar{\eta} (y) \delta \eta (z)} &= -
\int d^4 v d^4 w d^4 r \left( \frac{\delta^2 \Gamma}{\delta \bar{\psi} (z)
\delta \psi (v)} \right)^{- 1} \frac{\delta \phi (r)}{\delta J (x)} \left(
\frac{\delta^3 \Gamma}{\delta \phi (r) \delta \bar{\psi} (v) \delta \psi (w)}
\right) \left( \frac{\delta^2 \Gamma}{\delta \bar{\psi} (w) \delta \psi (y)}
\right)^{- 1} \nonumber \\
&= \int
d^4 v d^4 w d^4 r \left( \frac{\delta^2 \Gamma}{\delta \bar{\psi} (z) \delta
\psi (v)} \right)^{- 1} \left( \frac{\delta^2 W}{\delta J (x) \delta J (r)}
\right) \left( \frac{\delta^3 \Gamma}{\delta \phi (r) \delta \bar{\psi} (v)
\delta \psi (w)} \right) \left( \frac{\delta^2 \Gamma}{\delta \bar{\psi} (w)
\delta \psi (y)} \right)^{- 1}
\end{align}

Using the expression for the propagators ($\frac{\delta^2 W}{\delta J (z) \delta J (x)} = - i G_s (x - z)$ and $\frac{\delta^2 \Gamma}{\delta \bar{\psi} (y) \delta \psi (z)} = -i G^{- 1}_f (z - y)$) we may write:

\begin{align}
\frac{\delta^3 W}{\delta J (x) \delta \bar{\eta} (y) \delta \eta (z)} &= \int
d^4 v d^4 w d^4 r (i G_f (v - z)) (- i G_s (x - r)) \left( \frac{\delta^3
\Gamma}{\delta \phi (r) \delta \bar{\psi} (v) \delta \psi (w)} \right) (i G_f
(y - w)) \nonumber \\
& = i
\int d^4 v d^4 w d^4 r G_f (v - z) G_s (x - r) \left( \frac{\delta^3
\Gamma}{\delta \phi (r) \delta \bar{\psi} (v) \delta \psi (w)} \right) G_f (y - w)
\end{align}

Defining the fermion-scalar Yukawa vertex functional as
\begin{equation}
\frac{i \delta^3 \Gamma}{\delta \phi (r) \delta \bar{\psi} (v) \delta\psi (w)} = \Gamma^{(3)} (r, v, w)
\end{equation}

we finally obtain:
\begin{equation}
\label{threegam}
\frac{\delta^3 W}{\delta J (x) \delta \bar{\eta} (y) \delta \eta (z)} = \int
d^4 v d^4 w d^4 r G_f (v - z) G_s (x - r) \Gamma^{(3)} (r, v, w) G_f (y - w)
\end{equation}

\subsection{Pseudoscalar self-interaction Vertex $\Gamma^{(4)}$}

To get the dressed vertex that joins four scalar legs we derive the scalar propagator:

\begin{equation}
\frac{\delta^3 W}{\delta J (x) \delta J (y) \delta J (z)} = - i
\frac{\delta}{\delta J (x)} G_s (y - z) = 
\frac{\delta}{\delta J (x)} \left( \frac{\delta^2 \Gamma}{\delta \phi (z)
\delta \phi (y)} \right)^{- 1}
\end{equation}

We use that $\frac{d}{dx}M^{-1}(x)=-M^{-1}(\frac{d}{dx}M(x))M^{-1}(x)$ as before:

\begin{equation}
\frac{\delta^3 W}{\delta J (x) \delta J (y) \delta J (z)} = - \int d^4 v d^4 w
\left( \frac{\delta^2 \Gamma}{\delta \phi (z) \delta \phi (v)} \right)^{- 1}
\left( \frac{\delta^3 \Gamma}{\delta J (x) \delta \phi (v) \delta \phi (w)}
\right) \left( \frac{\delta^2 \Gamma}{\delta \phi (w) \delta \phi (y)}
\right)^{- 1}
\end{equation}

Taking the functional derivative again with respect to $J(t)$, we obtain after standard manipulations:

\begin{eqnarray}
\frac{\delta^4 W}{\delta J (t) \delta J (x) \delta J (y) \delta J (z)} &=&
\int d^4 v d^4 w d^4 v' d^4 w' \left( \frac{\delta^2 \Gamma}{\delta \phi (z)
\delta \phi (v')} \right)^{- 1} \left( \frac{\delta^3 \Gamma}{\delta J (t)
\delta \phi (v') \delta \phi (w')} \right) \nonumber\\
& & \times \left( \frac{\delta^2
\Gamma}{\delta \phi (w') \delta \phi (v)} \right)^{- 1} \left( \frac{\delta^3
\Gamma}{\delta J (x) \delta \phi (v) \delta \phi (w)} \right) \left(
\frac{\delta^2 \Gamma}{\delta \phi (w) \delta \phi (y)} \right)^{- 1} \nonumber\\
& & - \int d^4 v d^4 w \left( \frac{\delta^2 \Gamma}{\delta \phi (z) \delta \phi
(v)} \right)^{- 1} \left( \frac{\delta^4 \Gamma}{\delta J (t) \delta J (x)
\delta \phi (v) \delta \phi (w)} \right) \left( \frac{\delta^2 \Gamma}{\delta
\phi (w) \delta \phi (y)} \right)^{- 1} \nonumber\\
& & + \int d^4 v d^4 w d^4 v' d^4 w' \left( \frac{\delta^2 \Gamma}{\delta \phi
(z) \delta \phi (v)} \right)^{- 1} \left( \frac{\delta^3 \Gamma}{\delta J (x)
\delta \phi (v) \delta \phi (w)} \right) \left( \frac{\delta^2 \Gamma}{\delta
\phi (w) \delta \phi (v')} \right)^{- 1} \nonumber\\
& & \times\left( \frac{\delta^3 \Gamma}{\delta
J (t) \delta \phi (v') \delta \phi (w')} \right) \left( \frac{\delta^2
\Gamma}{\delta \phi (w') \delta \phi (y)} \right)^{- 1}
\end{eqnarray}

Using the chain rule of functional differentiation, and expressing the above quantity in terms of the propagators ($\frac{\delta^2 \Gamma}{\delta \phi (y) \delta \phi (z)} = i G^{- 1}_s (y -z)$ and $\frac{\delta^2 W}{\delta J (z) \delta J (x)} = - i G_s (x - z)$), we obtain:

\begin{eqnarray}
\frac{\delta^4 W}{\delta J (t) \delta J (x) \delta J (y) \delta J (z)} &=& - i
\int d^4 v d^4 w d^4 v' d^4 w' d^4 r d^4 r' G_s (v' - z) G_s (r - t) \left(
\frac{\delta^3 \Gamma}{\delta \phi (r) \delta \phi (v') \delta \phi (w')}
\right) \nonumber\\
& & \times G_s (v - w') G_s (r' - x) \left( \frac{\delta^3 \Gamma}{\delta \phi
(r') \delta \phi (v) \delta \phi (w)} \right) G_s (w - y) \nonumber\\
& & - \int d^4 v d^4 w d^4 r d^4 r' G_s (v - z) G_s (r - t) G_s (r' - x) \nonumber\\
& & \times \left(\frac{\delta^4 \Gamma}{\delta \phi (r) \delta \phi (r') \delta \phi (v) \delta
\phi (w)} \right) G_s (w - y)\nonumber\\
& & - i \int d^4 v d^4 w d^4 v' d^4 w' d^4 r d^4 r' G_s (v - z) G_s (r' - x)
\left( \frac{\delta^3 \Gamma}{\delta \phi (r') \delta \phi (v) \delta \phi
(w)} \right) \nonumber\\
& & \times G_s (v' - w) G_s (r - t) \left( \frac{\delta^3 \Gamma}{\delta
\phi (r) \delta \phi (v') \delta \phi (w')} \right) G_s (y - w')
\end{eqnarray}

However,  in our case,  $\frac{\delta^3 \Gamma}{\delta \phi (r') \delta \phi (v) \delta \phi (w)}
= 0$ since there is no interaction $\phi^3$, hence
 
\begin{eqnarray}
\frac{\delta^4 W}{\delta J (t) \delta J (x) \delta J (y) \delta J (z)} &=& - \int
d^4 v d^4 w d^4 r d^4 r' G_s (v - z) G_s (r - t) G_s (r' - x) \nonumber\\
& & \times \left(
\frac{\delta^4 \Gamma}{\delta \phi (r) \delta \phi (r') \delta \phi (v) \delta
\phi (w)} \right) G_s (w - y)
\end{eqnarray}

Defining, 
\begin{equation}
\frac{i \delta^4 \Gamma}{\delta \phi (r) \delta \phi (r') \delta \phi (v)
\delta \phi (w)} = \Gamma^{(4)} (r, r', v, w)\,.
\end{equation}
we finally obtain:

\begin{equation}
\frac{\delta^4 W}{\delta J (t) \delta J (x) \delta J (y) \delta J (z)} = i \int
d^4 v d^4 w d^4 r d^4 r' G_s (v - z) G_s (r - t) G_s (r' - x) \Gamma^{(4)} (r,
r', v, w) G_s (w - y)\,.
\end{equation}

\subsection{Schwinger-Dyson equations}

\subsubsection{Fermion equation}
We start from
\begin{equation}
  \int \mathcal{D} [\phi \psi \bar{\psi}] \frac{\delta}{\delta \bar{\psi}}
  e^{i S} = 0
\end{equation}

\begin{equation}
  \int \mathcal{D} [\phi \psi \bar{\psi}] \left( \eta (x) + i \slashed{\partial}
  \psi (x) - i g \phi (x) \gamma^5 \psi (x) \right) e^{i S} = 0
\end{equation}
with

\begin{align}\label{Sdef}
S = \int d^4 x' \left[ - \frac{1}{2} \phi \partial^2 \phi + \bar{\psi} i
\slashed{\partial} \psi - i g \phi \bar{\psi} \gamma^5 \psi + \frac{\lambda}{4!} \phi^4
\right] + \int d^4 x [J \phi + \bar{\psi} \eta + \bar{\eta} \psi].
\end{align}

We replace $\phi (x)$ and $\psi (x)$ with functional derivatives:
\begin{equation}
  \int \mathcal{D} [\phi \psi \bar{\psi}] \left( \eta (x) + i \slashed{\partial}
  \left( \frac{- i \delta}{\delta \bar{\eta} (x)} \right) - i g \left( \frac{-
  i \delta}{\delta J (x)} \right) \gamma^5 \left( \frac{- i \delta}{\delta
  \bar{\eta} (x)} \right) \right) e^{i S} = 0
\end{equation}

\begin{equation}
  \int \mathcal{D} [\phi \psi \bar{\psi}] \left( \eta (x) + \slashed{\partial}
  \left( \frac{\delta}{\delta \bar{\eta} (x)} \right) + i g \left(
  \frac{\delta}{\delta J (x)} \right) \gamma^5 \left( \frac{\delta}{\delta
  \bar{\eta} (x)} \right) \right) e^{i S} = 0
\end{equation}
from which
\begin{equation}
  \left[ \eta (x) + \slashed{\partial} \left( \frac{\delta}{\delta \bar{\eta} (x)}
  \right) + i g \left( \frac{\delta}{\delta J (x)} \right) \gamma^5 \left(
  \frac{\delta}{\delta \bar{\eta} (x)} \right) \right] Z = 0
\end{equation}
and using $Z=e^{iW}$, we finally obtain, after standard manipulations

\begin{equation}
  \eta (x) - i \slashed{\partial} \left( \frac{\delta W}{\delta \bar{\eta} (x)}
  \right) - i g \left( \frac{\delta W}{\delta J (x)} \right) \gamma^5 \left(
  \frac{\delta W}{\delta \bar{\eta} (x)} \right) + g \gamma^5 \left(
  \frac{\delta^2 W}{\delta J (x) \delta \bar{\eta} (x)} \right) = 0
\end{equation}
 Taking the functional derivative with respect to $\eta (y)$, 
 we obtain
 \begin{eqnarray}
\label{firstdereta}
  \delta^{(4)} (x - y) - i \slashed{\partial} \left( \frac{\delta^2 W}{\delta \eta
  (y) \delta \bar{\eta} (x)} \right) - i g \left( \frac{\delta^2 W}{\delta
  \eta (y) \delta J (x)} \right) \gamma^5 \left( \frac{\delta W}{\delta
  \bar{\eta} (x)} \right) \nonumber\\
+ i g \left( \frac{\delta W}{\delta J (x)} \right)
  \gamma^5 \left( \frac{\delta^2 W}{\delta \eta (y) \delta \bar{\eta} (x)}
  \right) - g \gamma^5 \left( \frac{\delta^3 W}{\delta J (x) \delta \eta (y)
  \delta \bar{\eta} (x)} \right) = 0
\end{eqnarray}
Setting the sources equal to zero, and using the propagators (definitions \eqref{propfer} and \eqref{threegam}), we finally arrive at:

\begin{equation}
  \delta^{(4)} (x - y) - \slashed{\partial} G_f (x - y) + g \gamma^5 \left( \int
  d^4 v d^4 w d^4 r G_f (x - v) G_s (x' - r) \Gamma^{(3)} (r, v, w) G_f (w -
  y) \delta (x - x') \right) = 0
\end{equation}
which can be manipulated to give:
\begin{align}\label{paso1}
  &  0 = \int d^4 y \delta^{(4)} (x - y) G^{- 1}_f (z - y) - \int d^4 y
  \slashed{\partial} G_f (x - y) G^{- 1}_f (z - y) \nonumber\\
&  + g \gamma^5 \left( \int d^4 y
  d^4 v d^4 w d^4 r G_f (x - v) G_s (x' - r) \Gamma^{(3)} (r, v, w) G_f (w -
  y) G^{- 1}_f (z - y) \delta (x - x') \right) = 0\nonumber\\
&= 
  G^{- 1}_f (x - z) - \slashed{\partial} \delta^{(4)} (x - z) + g \gamma^5 \left(
  \int d^4 v d^4 r G_f (x - v) \Gamma^{(3)} (r, v, z) G_s (x' - r) \delta (x -
  x') \right) \,,
\end{align}
and finally
\begin{equation}
  G^{- 1}_f (x - y) - S^{- 1}_f (x - y) + g \gamma^5 \left( \int d^4 v d^4 w
  G_f (x - v) \Gamma^{(3)} (w, v, y) G_s (x' - w) \delta (x - x') \right) = 0
\end{equation}

Upon going to the Fourier space, we write:
\begin{align}
 0 &=  G^{- 1}_f (x - y) - S^{- 1}_f (x - y) + \left( \int d^4 v d^4 w \int_p g
  \gamma^5 G_f (p) e^{- i p (x - v)} \right.\nonumber\\
&\left.\times\int_k \int_q \int_l \Gamma^{(3)} \left(
  q, l, k \right) e^{i q w} e^{- i l v} e^{i k y} \delta (l - q - k) \int_{p'}
  G_s (p') e^{i p' (x' - w)} \delta (x - x') \right) \nonumber \\
&=
  G^{- 1}_f (x - y) - S^{- 1}_f (x - y) + \left( \int d^4 v d^4 w \int_p
  \int_{p'} \int_k \int_q \int_l g \gamma^5 G_f (p) \Gamma^{(3)} \left( q, l,
  k \right) \right.\nonumber\\
&\left. \times G_s (p') \delta (l - q - k) e^{- i p x} e^{i p' x'} e^{- i (p' -
  q) w} e^{- i (l - p) v} e^{i k y} \delta (x - x') \right) \nonumber \\
 &= \int_k [G^{- 1}_f (k) - S^{- 1}_f (k)] e^{- i k (x - y)} + \left( \int_p
  \int_k g \gamma^5 G_f (p) \Gamma^{(3)} (p - k, p, k) G_s (p - k) e^{- i k (x
  - y)} \right) 
\end{align}
which implies the pseudoscalar SD equation
\begin{equation}
  G^{- 1}_f (k) - S^{- 1}_f (k) = - \int_p g \gamma^5 G_f (p)
  \Gamma^{(3)} (p, k) G_s (p - k)
\end{equation}
given schematically by the lower diagram of fig.~\ref{fig:fsprops}. 

\subsubsection{Pseudoscalar equation}

We start as before with
\begin{equation}
  \int \mathcal{D} [\phi \psi \bar{\psi}] \frac{\delta}{\delta \phi (x)} e^{i
  S} = 0
\end{equation}
with $S$ given by \eqref{Sdef}, which yields 
\begin{equation}
  \int \mathcal{D} [\phi \psi \bar{\psi}] \left( J (x) - \partial^2 \phi (x) -
  i g \bar{\psi} (x) \gamma^5 \psi (x) + \frac{\lambda}{3!} \phi (x)^3 \right) e^{i
  S} = 0\,.
\end{equation}
Using similar steps as for the fermion case, with $Z=e^{iW}$, we obtain after some standard calculations

\begin{eqnarray}\label{Jeq}
 & &  J (x) + \partial^2 \left( \frac{\delta W}{\delta J (x)} \right) + i g \left(
  \frac{\delta W}{\delta \eta (x)} \right) \gamma^5 \left( \frac{\delta
  W}{\delta \bar{\eta} (x)} \right) - g \frac{\delta}{\delta \eta (x)} \left(
  \gamma^5 \frac{\delta W}{\delta \bar{\eta} (x)} \right) - \frac{\lambda}{3!}
  \left( \frac{\delta W}{\delta J (x)} \right)^3 \nonumber\\
& & - i \frac{\lambda}{2} \left(
  \frac{\delta W}{\delta J (x)} \right) \left( \frac{\delta^2 W}{\delta J (x)
  \delta J (x)} \right) + \frac{\lambda}{3!} \left( \frac{\delta^3 W}{\delta J (x)
  \delta J (x) \delta J (x)} \right) = 0
\end{eqnarray}

By writing the fourth term  as

\begin{equation}
\frac{\delta}{\delta \eta (x)} \left( \gamma^5 \frac{\delta W}{\delta
\bar{\eta} (x)} \right) = \tmop{tr} \left[ \frac{\delta}{\delta \eta (x)}
\left( \gamma^5 \frac{\delta W}{\delta \bar{\eta} (x)} \right) \right] = -
\tmop{tr} \left[ \gamma^5 \frac{\delta^2 W}{\delta \eta (x) \delta \bar{\eta}
(x)} \right] = \tmop{tr} \left[ \gamma^5 \frac{\delta^2 W}{\delta \bar{\eta}
(x) \delta \eta (x)} \right]\,,
\end{equation}
substituting in \eqref{Jeq}, and 
deriving with respect to the source $J (y)$, we obtain:
\begin{eqnarray}
\label{firstderj}
& &  0 = \delta^{(4)} (x - y) + \partial^2 \left( \frac{\delta^2 W}{\delta J (y)
  \delta J (x)} \right) + i g \left( \frac{\delta^2 W}{\delta J (y) \delta
  \eta (x)} \right) \gamma^5 \left( \frac{\delta W}{\delta \bar{\eta} (x)}
  \right) \nonumber\\
& & + i g \left( \frac{\delta W}{\delta \eta (x)} \right) \gamma^5
  \left( \frac{\delta^2 W}{\delta J (y) \delta \bar{\eta} (x)} \right) - g
  \tmop{tr} \left[ \gamma^5 \frac{\delta^3 W}{\delta J (y) \delta \bar{\eta}
  (x) \delta \eta (x)} \right] \nonumber\\
& & - \frac{\lambda}{2} \left( \frac{\delta W}{\delta J
  (x)} \right)^2 \left( \frac{\delta^2 W}{\delta J (y) \delta J (x)} \right) -
  i \frac{\lambda}{2} \left( \frac{\delta^2 W}{\delta J (y) \delta J (x)} \right)
  \left( \frac{\delta^2 W}{\delta J (x) \delta J (x)} \right) \nonumber\\
& & - i \frac{\lambda}{2}
  \left( \frac{\delta W}{\delta J (x)} \right) \left( \frac{\delta^3 W}{\delta
  J (y) \delta J (x) \delta J (x)} \right) + \frac{\lambda}{3!} \left(
  \frac{\delta^4 W}{\delta J (y) \delta J (x) \delta J (x) \delta J (x)}
  \right) \nonumber\\
\end{eqnarray}
Setting $\delta W/\delta (J (x), \bar \eta(x), \eta) =0$,
we obtain
\begin{eqnarray}
 & &  \delta^{(4)} (x - y) + \partial^2 \left( \frac{\delta^2 W}{\delta J (y)
  \delta J (x)} \right) - g \tmop{tr} \left[ \gamma^5 \frac{\delta^3 W}{\delta
  J (y) \delta \bar{\eta} (x) \delta \eta (x)} \right] \nonumber\\
& & - i \frac{\lambda}{2} \left(
  \frac{\delta^2 W}{\delta J (y) \delta J (x)} \right) \left( \frac{\delta^2
  W}{\delta J (x) \delta J (x)} \right) + \frac{\lambda}{3!} \left( \frac{\delta^4
  W}{\delta J (y) \delta J (x) \delta J (x) \delta J (x)} \right) = 0\nonumber\\
\end{eqnarray}
Writing in terms of the vertices and propagators, we eventually obtain, after some tedious but standard manipulations, similar to the fermion case:
\begin{eqnarray}
\label{paso2}
& &  G^{- 1}_s (x - z) - i \partial^2 \delta^{(4)} (x - z) - \tmop{tr} \left[ g
  \gamma^5 \int d^4 v d^4 w G_f (x - v) \Gamma^{(3)} (z, v, w) G_f (w - x)
  \right]  \nonumber\\
& & + i \frac{\lambda}{2} \delta^{(4)} (x - z) G_s (x - x)  + i \frac{\lambda}{3!} \int d^4 v d^4 w d^4 r G_s (v - x) G_s (r - x) \Gamma^{(4)}
(z, r, v, w) G_s (w - x)= 0
\end{eqnarray}

 Therefore,
\begin{eqnarray}
& &   G^{- 1}_s (x - y) - S^{- 1}_s (x - y) - \tmop{tr} \left[ g \gamma^5 \int d^4
  v d^4 w G_f (x - v) \Gamma^{(3)} (y, v, w) G_f (w - x) \right] \nonumber\\
& & + i  \frac{\lambda}{2} \delta^{(4)} (x - y) G_s (x - x) + i \frac{\lambda}{3!} \int d^4 v d^4 w d^4 r G_s (v - x) G_s (r - x) \Gamma^{(4)}
(y, r, v, w) G_s (w - x) = 0 \nonumber\\
\end{eqnarray}

Going to Fourier space, this yields, after some calculations, similar to the fermion case:

\begin{eqnarray}
  G^{- 1}_s (k) - S^{- 1}_s (k) &=& \tmop{tr} \left[ \int_p g \gamma^5 G_f (p)
  \Gamma^{(3)} (p, k) G_f (p - k) \right] - i \frac{\lambda}{2} \int_p G_s (p)  \nonumber\\
& &  - i \frac{\lambda}{3!} \int_p \int_l G_s (p) G_s (k + l - p) \Gamma^{(4)} (k, p, l) G_s (l) \,,
\end{eqnarray}
which is depicted schematically in the upper diagram of fig,~\ref{fig:fsprops}.

\subsubsection{Vertex $\phi\bar{\psi}\psi$}

We start from \eqref{firstdereta} by taking the functional derivative with respect to the source $J(z)$:
\begin{eqnarray}
& & - i \slashed{\partial} \left( \frac{\delta^3 W}{\delta J (z) \delta \eta (y)
\delta \bar{\eta} (x)} \right) - i g \left( \frac{\delta^3 W}{\delta J (z)
\delta \eta (y) \delta J (x)} \right) \gamma^5 \left( \frac{\delta W}{\delta
\bar{\eta} (x)} \right) - i g \left( \frac{\delta^2 W}{\delta \eta (y) \delta
J (x)} \right) \gamma^5 \left( \frac{\delta^2 W}{\delta J (z) \delta
\bar{\eta} (x)} \right)\nonumber\\
& & + i g \left( \frac{\delta^2 W}{\delta J (z) \delta J (x)} \right) \gamma^5
\left( \frac{\delta^2 W}{\delta \eta (y) \delta \bar{\eta} (x)} \right) + i g
\left( \frac{\delta W}{\delta J (x)} \right) \gamma^5 \left( \frac{\delta^3
W}{\delta J (z) \delta \eta (y) \delta \bar{\eta} (x)} \right) \nonumber\\
& & - g \gamma^5\left( \frac{\delta^4 W}{\delta J (z) \delta J (x) \delta \eta (y) \delta
\bar{\eta} (x)} \right) = 0
\end{eqnarray}

Setting the sources and the one point function to zero yields, and expressing the resulting expression 
in terms of the vertices and propagators:

\begin{eqnarray}
& & \int d^4 x' d^4 v d^4 w d^4 r i \slashed{\partial} \delta^{(4)} (x - x') G_f
(x' - v) G_s (z - r) \Gamma^{(3)} (r, v, w) G_f (w - y) 
\nonumber\\ 
& & - i g (- i G_s (x -
z)) \gamma^5 (i G_f (x - y)) = 0
\end{eqnarray}

To eliminate the partial derivative we use the last equality in \eqref{paso1}:
\begin{eqnarray}
& & \int d^4 x' d^4 v d^4 w d^4 r \left[ G^{- 1}_f (x - x') + g \gamma^5 \left(
\int d^4 v d^4 r G_f (x - v) \Gamma^{(3)} (r, v, x') G_s (x'' - r) \delta (x -
x'') \right) \right] \nonumber\\
& & \times G_f (x' - v) G_s (z - r) \Gamma^{(3)} (r, v, w) G_f (w -
y) - g G_s (x - z) \gamma^5 G_f (x - y) = 0
\end{eqnarray}
Multiplying by $G_f (s - x)$  we get:
\begin{eqnarray}
& & \int d^4 w d^4 r G_s (z - r) \Gamma^{(3)} (r, s, w) G_f (w - y) G_f (s - x)= \nonumber\\
& & g G_s (s - z) \gamma^5 G_f (s - y) G_f (s - x)\nonumber\\
& & - \int d^4 x' d^4 v d^4 w d^4 r \left[ g \gamma^5 \left( \int d^4 v' d^4 r'
G_f (s - v') \Gamma^{(3)} (r', v', x') G_s (x'' - r') \delta (s - x'') \right)
\right] \nonumber\\
& & \times G_f (x' - v) G_s (z - r) \Gamma^{(3)} (r, v, w) G_f (w - y) G_f (s -x) 
\end{eqnarray}

Going to Fourier space, we finally obtain :
\begin{equation}
\Gamma^{(3)} (q, p, p') = g \gamma^5 - \int_k [g \gamma^5 G_f (k)
\Gamma^{(3)} (k, p) G_s (p - k)] G_f (q - p') \Gamma^{(3)} (q, q - p', p')\,,
\end{equation}
which is given schematically by the upper diagram of fig.~\ref{fig:vert}. 

\subsubsection{Vertex $\phi^4$}

We start from \eqref{firstderj}, deriving it functionally twice with respect to $J(z)$ and $J(t)$.
Setting the sources and the one point function to zero, we get:
\begin{eqnarray}
& & \partial^2 \left( \frac{\delta^4 W}{\delta J (t) \delta J (z) \delta J (y)
\delta J (x)} \right) - u \left( \frac{\delta^2 W}{\delta J (t) \delta J (x)} \right) \left(
\frac{\delta^2 W}{\delta J (z) \delta J (x)} \right) \left( \frac{\delta^2
W}{\delta J (y) \delta J (x)} \right)\nonumber\\
& & - i \frac{\lambda}{2} \left( \frac{\delta^4 W}{\delta J (t) \delta J (z) \delta J
(y) \delta J (x)} \right) \left( \frac{\delta^2 W}{\delta J (x) \delta J (x)}
\right) - i \frac{\lambda}{2} \left( \frac{\delta^2 W}{\delta J (y) \delta J (x)} \right)
\left( \frac{\delta^4 W}{\delta J (t) \delta J (z) \delta J (x) \delta J (x)}
\right)\nonumber\\
& & - i \frac{\lambda}{2} \left( \frac{\delta^2 W}{\delta J (z) \delta J (x)} \right)
\left( \frac{\delta^4 W}{\delta J (t) \delta J (y) \delta J (x) \delta J (x)}
\right) - i \frac{\lambda}{2} \left( \frac{\delta^2 W}{\delta J (t) \delta J (x)} \right)
\left( \frac{\delta^4 W}{\delta J (z) \delta J (y) \delta J (x) \delta J (x)}
\right) = 0\nonumber\\
\end{eqnarray}

Introducing a delta function, multiplying by $i$ and using the propagators and vertices we obtain:
\begin{eqnarray}
& & i \int d^4 x' i \partial^2 \delta (x - x') \int d^4 v d^4 w d^4 r d^4 r' G_s (v
- x') G_s (r - t) G_s (r' - z) \Gamma^{(4)} (r, r', v, w) G_s (w - y)\nonumber\\
& & - i \lambda (- i G_s (x - t)) (- i G_s (x - z)) (- i G_s (x - y))\nonumber\\
& & + i \frac{\lambda}{2} \int d^4 v d^4 w d^4 r d^4 r' G_s (v - x) G_s (r - t) G_s (r' -
z) \Gamma^{(4)} (r, r', v, w) G_s (w - y) (- i G_s (x - x))\nonumber\\
& & + i \frac{\lambda}{2} (- i G_s (x - y)) \int d^4 v d^4 w d^4 r d^4 r' G_s (v - x) G_s
(r - t) G_s (r' - z) \Gamma^{(4)} (r, r', v, w) G_s (w - x)\nonumber\\
& & + i \frac{\lambda}{2} (- i G_s (x - z)) \int d^4 v d^4 w d^4 r d^4 r' G_s (v - x) G_s
(r - t) G_s (r' - y) \Gamma^{(4)} (r, r', v, w) G_s (w - x)\nonumber\\
& & + i \frac{\lambda}{2} (- i G_s (x - t)) \int d^4 v d^4 w d^4 r d^4 r' G_s (v - x) G_s
(r - z) G_s (r' - y) \Gamma^{(4)} (r, r', v, w) G_s (w - x) = 0\nonumber\\
\end{eqnarray}

Then, on using equation \eqref{paso2} to eliminate the partial derivative, and 
multiplying by $G_s (s - x)$, we obtain after standard manipulations:
\begin{eqnarray}
& & i \int d^4 w d^4 r d^4 r' G_s (r - t) G_s (r' - z) \Gamma^{(4)} (r, r', s, w)
G_s (w - y) G_s (s - x) =\nonumber\\
& & - \lambda\, G_s (s - t) G_s (s - z) G_s (s - y) G_s (s - x)\nonumber\\
& & - \frac{\lambda}{2} \int d^4 v d^4 w d^4 r d^4 r' G_s (v - s) G_s (r - t) G_s (r'
- z) \Gamma^{(4)} (r, r', v, w) G_s (w - s) G_s (s - x) G_s (s - y)\nonumber\\
& & - \frac{\lambda}{2}  \int d^4 v d^4 w d^4 r d^4 r' G_s (v - s) G_s (r - t) G_s
(r' - y) \Gamma^{(4)} (r, r', v, w) G_s (w - s) G_s (s - z) G_s (s - x)\nonumber\\
& & - \frac{\lambda}{2} \int d^4 v d^4 w d^4 r d^4 r' G_s (v - s) G_s (r - z) G_s (r'
- y) \Gamma^{(4)} (r, r', v, w) G_s (w - s) G_s (s - t) G_s (s - x)\nonumber\\
& & + \frac{\lambda}{3!} \int d^4 x' d^4 v d^4 w d^4 r d^4 r' \int d^4 v' d^4 w' d^4
r'' G_s (v' - s) G_s (r'' - s) \Gamma^{(4)} (x', r'', v', w') \nonumber\\
& & \times G_s (w' - s) G_s
(v - x') G_s (r - t) G_s (r' - z) \Gamma^{(4)} (r, r', v, w) G_s (w - y) G_s(s - x)\nonumber\\
& & + i \int d^4 x' d^4 v d^4 w d^4 r d^4 r' \left[ \tmop{tr} \left[ g \gamma^5
\int d^4 v' d^4 w' G_f (s - v') \Gamma^{(3)} (x', v', w') G_f (w' - s) \right]\right] \nonumber\\
& & \times G_s (v - x') G_s (r - t) G_s (r' - z) \Gamma^{(4)} (r, r', v, w) G_s
(w - y) G_s (s - x)\nonumber\\
\end{eqnarray}

Going to Fourier space, we finally obtain:
\begin{eqnarray}
\Gamma^{(4)} (q, q', p, p') &=& i \lambda \nonumber\\
& & + i \frac{\lambda}{2} \int_k G_s (q + q' - k) \Gamma^{(4)} (q, q', k, q + q' - k)G_s (k)\nonumber\\
& & + i \frac{\lambda}{2}  \int_k G_s (k) \Gamma^{(4)} (k, q', k + p - q, p') G_s (k +p - q)\nonumber\\
& & + i \frac{\lambda}{2} \int_k G_s (k) \Gamma^{(4)} (q, q' + k - p, k, p') G_s (q' +k - p)\nonumber\\
& & - i \frac{\lambda}{3!} \int_k \int_{k'} G_s (k) G_s (k') \Gamma^{(4)} (k + k' - p, q
+ q' - p', k, k') G_s (k + k' - p) \nonumber\\
& & \times G_s (q + q' - p') \Gamma^{(4)} (q, q', q + q' - p', p')\nonumber\\
& & + \int_k \tmop{tr} [g \gamma^5 G_f (k) \Gamma^{(3)} (k, p) G_f (k - p)] G_s
(q + q' - p') \Gamma^{(4)} (q, q', q + q' - p', p')\,,
\end{eqnarray}
which is depicted schematically in the lower diagram of fig.~\ref{fig:vert}.

\end{document}